\documentclass[12pt,a4paper]{article}
\usepackage[utf8]{inputenc}
\usepackage[english]{babel}
\usepackage{amsmath}
\usepackage{subfigure}
\usepackage{amsfonts}
\usepackage{amssymb}
\usepackage{graphicx}
\newcommand{\be}{\begin{equation}}
\newcommand{\ee}{\end{equation}}
\newcommand{\bea}{\begin{eqnarray}}
\newcommand{\eea}{\end{eqnarray}}

\newcommand{\gbl}{g_{\rm BL}}
\newcommand{\vbl}{v_{\rm BL}}
\newcommand{\zbl}{Z_{\rm BL}}
\newcommand{\mzbl}{M_{\zbl}}
\newcommand{\mdm}{M_{N_1}}
\newcommand{\ubl}{{\rm U}(1)_{\rm B-L}}
\newcommand{\bl}{{\rm B-L}}

\catcode`@=12
\def\la{\mathrel{\mathchoice {\vcenter{\offinterlineskip\halign{\hfil
$\displaystyle##$\hfil\cr<\cr\sim\cr}}}
{\vcenter{\offinterlineskip\halign{\hfil$\textstyle##$\hfil\cr<\cr\sim\cr}}}
{\vcenter{\offinterlineskip\halign{\hfil$\scriptstyle##$\hfil\cr<\cr\sim\cr}}}
{\vcenter{\offinterlineskip\halign{\hfil$\scriptscriptstyle##$\hfil\cr<\cr\sim
\cr}}}}}
\def\ga{\mathrel{\mathchoice {\vcenter{\offinterlineskip\halign{\hfil
$\displaystyle##$\hfil\cr>\cr\sim\cr}}}
{\vcenter{\offinterlineskip\halign{\hfil$\textstyle##$\hfil\cr>\cr\sim\cr}}}
{\vcenter{\offinterlineskip\halign{\hfil$\scriptstyle##$\hfil\cr>\cr\sim\cr}}}
{\vcenter{\offinterlineskip\halign{\hfil$\scriptscriptstyle##$\hfil\cr>\cr\sim
\cr}}}}}
\begin{document}
\thispagestyle{empty}
\begin{center}
{\Large \bf
{Freeze-in Production of Sterile Neutrino Dark Matter in $\mathbf{\ubl}$ Model}}\\
\vspace{0.25cm}
\begin{center}
{{\bf Anirban Biswas} \footnote[1]{\makebox[1.cm]{Email:} anirbanbiswas@hri.res.in},
{\bf Aritra Gupta}\footnote[2]{\makebox[1.cm]{Email:} aritra@hri.res.in}}\\
\vspace{0.5cm}
{\it Harish-Chandra Research Institute, Chhatnag Road,
Jhunsi, Allahabad 211 019, INDIA}
\end{center}
\vspace{1cm}
{\bf ABSTRACT} \\
\end{center}
With the advent of new and more sensitive direct detection
experiments, scope for a thermal WIMP explanation of dark matter
(DM) has become extremely constricted. The non-observation of
thermal WIMP in these experiments has put a strong upper bound on
WIMP-nucleon scattering cross section and within a few years it is
likely to overlap with the coherent neutrino-nucleon
cross section. Hence in all probability, DM may
have some non-thermal origin. In this work we explore in detail this
possibility of a non-thermal sterile neutrino DM within the framework
of U(1)$_{\rm B-L}$ model. The U(1)$_{\rm B-L}$ model on the other hand
is a well-motivated and minimal way of extending the standard model so
that it can explain the neutrino masses via Type-I see-saw mechanism.
We have shown, besides explaining the neutrino mass, it can also
accommodate a non-thermal sterile neutrino DM with correct relic density.
In contrast with the existing literature, we have found that $W^\pm$ decay
can also be a dominant production mode of the sterile neutrino DM. To obtain
the comoving number density of dark matter, we have solved here a coupled set
of Boltzmann equations considering all possible decay as well as annihilation
production modes of the sterile neutrino dark matter. The framework developed
here though has been done for a U(1)$_{\rm B-L}$ model, can be applied quite
generally for any models with an extra neutral gauge boson and a fermionic
non-thermal dark matter.
\vskip 1cm
\newpage
\section{Introduction}
\label{intro}
The existence of Dark Matter (DM) in the Universe is now an
acceptable reality. There are various satellite borne experiments,
namely WMAP \cite{1212.5226} and Planck \cite{1502.01589} who have already
measured the current mass density (relic density) of DM in the
Universe with an extremely good accuracy. Moreover, there are also some
indirect evidences about the existence of dark matter such as flatness
of galactic rotation curve \cite{astro-ph/0010594}, gravitational lensing of distant
object \cite{astro-ph/9912508}, bullet cluster
\cite{astro-ph/0312273} etc. However the composition
of DM is still unknown to us. The Standard Model (SM) of electroweak
interaction does not have any fundamental particle which can play
the role of DM. Hence in order to accommodate a viable dark matter
candidate we need to formulate a theory beyond Standard Model (BSM)
of electroweak interaction. Among the various possible BSM theories
available in literature the Weakly Interacting Massive Particle
(WIMP) is the most favourable class of dark matter candidates
and until now neutralino in the Supersymmetric Standard Model
is one of the most studied WIMPs \cite{hep-ph/9506380}. The presence of DM
is also being investigated in various direct detection experiments, namely
LUX \cite{1512.03506}, XENON 100 \cite{1207.5988} etc, and no ``real signal'' due to
a dark matter particle has been observed yet. With the increasing
sensitivity of the direct detection experiments
(``ton-scale") \cite{Aprile:2015uzo, 1509.02910,
Aalseth:2015mba}, the WIMP-nucleon
cross section is soon to merge with the elastic
neutrino-nucleon cross section \cite{Drukier:1983gj, 1310.8327}. The floor
mostly comprises of  8$_{\rm B}$ and 7$_{\rm Be}$
solar neutrinos \cite{1208.5723}. Hence in future our only probe to
distinguish a dark matter signal (assuming that DM is a thermal WIMP)
from the neutrino background will be through directional searches \cite{1602.03781}.
But if we wish to move beyond this thermal WIMP scenario, there is another class of
dark matter candidates which are produced through non-thermal
processes at an early stage of the Universe. 
Such possibilities include axino \cite{hep-ph/9905212, hep-ph/0101009, 1307.3330},
gravitino \cite{hep-th/9907124, hep-ph/9911302},
very heavy dark matter candidates like WIMPzillas \cite{hep-ph/9810361}
among many others \cite{1407.0017}. Their interaction
strengths with other particles (in the thermal plasma) are so
feeble that they never attain thermal equilibrium. These types
of dark matter candidates are known as Feebly Interacting Massive
Particle or FIMP \cite{0911.1120}. In contrast with the commonly discussed
WIMP scenario, the relic density of FIMP type dark matter
is attained by the so called Freeze-in mechanism \cite{0911.1120}.
Unlike the thermal Freeze-out mechanism where relic density
depends on the final abundance of dark matter, in Freeze-in,
DM relic density is sensitive to its initial
production history (for a nice review see \cite{1512.02751}).
In literature two types of Freeze-in mechanisms
are usually discussed, IR (infra-red)
Freeze-in \cite{hep-ph/0106249, 1105.1654, 1501.02666}
and UV (ultra-violet) Freeze-in
\cite{1410.6157, 1412.4791, Dev:2013yza}. Unlike the former,
the DM relic density in UV Freeze-in depends explicitly
on the reheat temperature (T$_{\rm R}$).
Production of the non-thermal DM candidate usually
occurs via a decay of a heavy mother particle
(e.g. from Inflaton decay and decay of heavy Moduli
fields \cite{hep-ph/0606075, hep-ph/0604236}).

In this work we will study a FIMP type dark matter candidate
in the U(1)$_{\rm B-L}$ extension of the Standard Model of
particle physics. U(1)$_{\rm B-L}$ extension of SM is a
very well motivated BSM theory as it provides the explanation
of nonzero neutrino mass through Type-I sea-saw mechanism. In this
model besides the usual SM gauge (${\rm SU}(3)_{\rm c} \times{\rm SU}(2)_{\rm L}
\times {\rm U}(1)_{\rm Y}$) symmetry, an additional local U(1)$_{\rm B-L}$
symmetry invariance is also imposed on the Lagrangian where $B$ and $L$
respectively represent the baryon and lepton number of a particle. In order
to obtain an ``anomaly free gauge theory'', three additional right
handed neutrinos ($N_i$, $i=1,\,3$) are required to be added
to the particle spectrum of SM. Moreover, we also require a complex scalar
($\Psi$) which is a singlet under the SM gauge group but possesses
a suitable nonzero U(1)$_{\rm B-L}$ charge. Majorana masses for the
three right handed neutrinos are generated through the spontaneous
breaking of the local ${\rm B-L}$ symmetry by the vacuum expectation value
(VEV) of complex scalar singlet $\Psi$. The lightest one ($N_1$) among
the three right handed neutrinos can be a viable
dark matter candidate.

The dark matter candidate $N_1$ in $\ubl$ model can be produced through
both thermal as well as non-thermal processes.
In the former case, the interaction strengths of DM particles
with others in the early Universe are such that they are able
to maintain their thermal as well as chemical equilibrium.
The decoupling of the DM particles occur when their interaction
rates fall short of the expansion rate of the Universe.
If $n_{\rm eq}$ and $\langle\sigma {\rm v}\rangle$ are the equilibrium
number density and the thermally averaged annihilation cross section
of $N_1$ then the decoupling condition requires
$\frac{n_{\rm eq}\,\langle\sigma {\rm v}\rangle}{H}<1$
with $H$ being the Hubble parameter. Being out of equilibrium,
the relic density of $N_1$ freezes to a particular value
which depends upon the interaction strength as well as the
temperature of the Universe at which the decoupling occurred (freeze-out
temperature). The thermally produced $N_1$ as a dark matter candidate,
in the U(1)$_{\rm B-L}$ extension of SM, has been studied in
Refs.\cite{1002.2525, 1308.0023, 1509.04036, 1604.06566}.
In these works most of the authors have shown that
the relic abundance of dark matter particle satisfied the WMAP or Planck limit
only when the mass of DM is nearly half the masses of mediating scalar
particles (at or near resonances). This requires significant
fine tuning as there is no symmetry, in the Lagrangian, which
can relate the masses of dark matter and the scalar sector
particles in the above mentioned way. Hence, with respect
to the above discussions, it is natural to think about a
dark matter particle, in this U(1)$_{\rm B-L}$ model, which
is produced through some non-thermal interactions at the early stage of the
Universe. Non-thermal sterile neutrino production from the
oscillation of active neutrinos was first proposed
by Dodelson-Widrow \cite{hep-ph/9303287},
but this idea is now in conflict with the X-ray observations \cite{1309.4091}.
Other mechanisms of sterile neutrino production
like Shi-Fuller mechanism \cite{astro-ph/9810076}
can alleviate some of these problems producing a colder dark matter spectrum.
Several other models have also successfully discussed
non-thermal sterile neutrino dark matter. They include some Supersymmetric
models \cite{1506.08195}, models using warped
extra-dimensions \cite{0711.1570} and decay
from charged \cite{1409.0659} and neutral
scalars \cite{1306.3996, 1502.01011} or from extra
gauge bosons \cite{1403.2727, 1606.09317}.
Most of the studies involving production of
sterile neutrino from extra gauge boson assume the gauge boson to
be in thermal equilibrium with the other SM particles. However, in
this work we have moved way from this assumption (details later).
Several non-thermal models of sterile neutrino dark
matter under the assumption of low reheating temperature have
also been studied in \cite{astro-ph/0403323, 0803.2735, 0804.0336}
which is also not the case we are considering here.

Additionally, unlike what is usually done in building a dark matter model,
we do not impose any extra symmetry to stabilise our dark matter candidate.
For an $\mathcal{O}$ (MeV) sterile neutrino dark matter we have a
dominant decay mode to $e^\pm$ and $\nu$ with a very large life time
(larger than the age of the Universe for the parameters we consider here)
which in turn helps us to propose a possible indirect detection
signal of the 511 keV line observed by INTEGRAL/SPI \cite{astro-ph/0309442} of ESA.  

The rest of the paper is organised as follows: In Section \ref{model}
we briefly describe the $\ubl$ model. In Section \ref{NTR} we describe
the production mechanism of non-thermal sterile neutrino dark matter
in detail. Section \ref{sec_BE} describes the Boltzmann equation(s)
needed to compute the comoving number densities of both $\zbl$ and $N_1$.
Calculation of relic density of sterile neutrino dark matter is given in
Section \ref{sec_RD}. Section \ref{integral} deals with a possible
indirect detection mode of our dark matter particle $N_1$. Finally
our conclusion is given in Section 7. All analytic expressions
of decay widths and annihilation cross sections used in
this work are listed in the Appendix.
\section{The U(1)$_{\rm B-L}$ extension of Standard Model}
\label{model}
In the present work we have considered a {\it minimal} $\ubl$ extension of the
Standard Model where the SM gauge sector is enhanced by
an additional local U(1)$_{\rm B-L}$ gauge symmetry with B and L are
known as the baryon and lepton number of a particle. Therefore, under the
$\ubl$ gauge group all SM leptons (including neutrinos) and
quarks have charges $-1$ and $\frac{1}{3}$ respectively. Besides the
SM fields, this model requires the presence of three right handed neutrinos
($N_i$, $i=1$, to 3) with $\ubl$ charge $-1$ for anomaly cancellation. On the
other hand, as the SM Higgs doublet ($\Phi$) does not possess any $\bl$
charge, hence in order to spontaneously break the local $\bl$ symmetry 
one needs to introduce a scalar field which transforms nontrivially under the $\ubl$
symmetry group. As a result, the scalar sector of the present model
is composed of a usual Higgs doublet (doublet under SU(2)$_{\rm L}$)
$\Phi$ and a complex scalar singlet $\Psi$. To generate
Majorana mass terms in a gauge invariant manner for the three right handed
neutrinos one needs the $\bl$ charge of $\Psi$ is $+2$. $\bl$
symmetry is spontaneously broken when $\Psi$ acquires VEV $\vbl$
while the remnant electroweak symmetry (SU(2)$_{\rm L}\times$ U(1)$_{\rm Y}$)
of the Lagrangian breaks spontaneously through the usual Higgs mechanism.
In unitary gauge, the expressions of $\Phi$ and $\Psi$, after getting VEVs
$v$ and $\vbl$ respectively, are 
\begin{eqnarray}
\Phi = \left(\begin{array}{c} 0 \\\frac{\phi + v}{\sqrt{2}}\end{array}\right)
\,,\,\,\,\,\,\,\,\,\,\,\,\,\Psi = \frac{\psi + v_{\rm BL}}{\sqrt{2}}\,.
\end{eqnarray} 
The gauge invariant and renormalisable Lagrangian of the
scalar sector is thus given by
\begin{eqnarray}
\mathcal{L}_{\rm scalar} &=& ({D_{\phi}}_{\mu} \Phi)^\dagger
({D_{\phi}}^{\mu} \Phi) + ({D_{\psi}}_{\mu} \Psi)^\dagger
({D_{\psi}}^{\mu} \Psi) -~V(\Phi, \Psi)\,\,\,,
\end{eqnarray}
with
\begin{eqnarray}
V(\Phi, \Psi) &=& \mu^2_1 (\Phi^\dagger\Phi) + \lambda_1 (\Phi^\dagger\Phi)^2 +
\mu^2_2 (\Psi^\dagger\Psi) + \lambda_2 (\Psi^\dagger\Psi)^2 \nonumber \\
&&
+\lambda_3 (\Phi^\dagger\Phi)(\Psi^\dagger\Psi)\,,
\label{scalar-potential}
\end{eqnarray}
where 
\begin{eqnarray}
{D_{\phi}}_{\mu}\Phi &=& \left(\partial_{\mu} + i \frac{g}{2}
\sigma^a {W_a}_{\mu } + i \frac{g^\prime}{2} B_{\mu} \right)\Phi \ , \nonumber \\
{D_{\psi}}_{\mu} \Psi &=& \left(\partial_{\mu} +
i\,Q_{\rm BL}(\Psi)\,{g_{\rm BL}}\,{Z_{\rm BL}}_{\mu} \right)\Psi \ ,
\end{eqnarray}
are the covariant derivatives of the scalar doublet $\Phi$ and
complex scalar singlet $\Psi$ respectively while $Q_{\rm BL}(\Psi)= +2$
is the ${\rm B-L}$ charge of $\Psi$. Gauge couplings of
SU(2)$_{\rm L}$, U(1)$_{\rm Y}$ and $\ubl$ are denoted
by $g$, $g^{\prime}$ and $\gbl$. The corresponding gauge
fields are $W_{a\,\mu}$ ($a=1$, 2, $3$), $B_{\mu}$ and ${\zbl}_{\mu}$.
After spontaneous
breaking of ${\rm SU}(2)_{\rm L}\times {\rm U}(1)_{\rm Y}\times {\rm U}(1)_{\rm B-L}$
symmetry by the VEVs of $\Phi$ and $\Psi$ we get two physical
neutral scalar fields $h$ and $H$ which can be expressed as a
linear combinations of $\phi$ and $\psi$ in the following way
\begin{eqnarray}
\left(\begin{array}{c} h \\ H\end{array}\right)
=\left(\begin{array}{cc}\cos\theta ~-\sin\theta
\\ \sin\theta ~~~~\cos\theta \end{array}\right)
\left(\begin{array}{c} \phi \\ \psi\end{array}\right) \,\, ,
\label{hH-phipsi}
\end{eqnarray} 
where $\theta$ is the mixing angle between the neutral
scalars $h$ and $H$. The expressions of mixing angle
($\theta$) and masses ($M_h$, $M_H$) of $h$ and $H$ are given by
\begin{eqnarray}
\theta &=& \frac{1}{2}\tan^{-1}\Bigg(\frac{\lambda_3\,\vbl\,v}
{\lambda_2 v^2_{\rm BL}-\lambda_1 v^2}\Bigg)\,, \nonumber \\
M^2_{h} &=& \lambda_1 v^2 + \lambda_2 v^2_{\rm BL} - 
\sqrt{(\lambda_1 v^2 - \lambda_2\,v^2_{\rm BL})^2 +
(\lambda_3 v\,v_{\rm BL})^2} \ ,\nonumber \\
M^2_{H} &=& \lambda_1 v^2 + \lambda_2 v^2_{\rm BL} + 
\sqrt{(\lambda_1 v^2 - \lambda_2 v^2_{\rm BL})^2
+ (\lambda_3 v v_{\rm BL})^2}  \,.
\label{scalar_mass}
\end{eqnarray}
We have considered the physical scalar $h$ as the SM-like Higgs boson
which was discovered recently by ATLAS \cite{1207.7214} and CMS \cite{1207.7235}
collaborations and consequently we have fixed the value of $M_h$
at 125.5 GeV. Also according to the measured values of Higgs boson
signal strengths (for its various decay modes) the mixing angle $\theta$ between
the SM-like Higgs boson $h$ and extra scalar boson $H$
should be very small. As this mixing angle does not play
any significant role in the present context, we have
kept $\theta$ fixed at 0.1 rad \cite{Basak:2014sza},
throughout this work, such that it satisfies all results
from both ATLAS and CMS collaborations. Besides this,
in order to obtained a stable vacuum the quatic couplings
of the Lagrangian (Eq. \ref{scalar-potential}) must satisfy
the following inequalities,
\begin{eqnarray}
&&\lambda_1 \geq 0\,,\nonumber\\
&&\lambda_2 \geq 0\,,\nonumber \\
&&\lambda_3 \geq -2\,\sqrt{\lambda_1\,\lambda_2}\,.
\end{eqnarray}
Moreover, as both $\Phi$ and $\Psi$ have nonzero VEVs, this
requires $\mu_i^2<0$ ($i=1,\,2$).

The gauge sector Lagrangian of the present model is given as
\footnote{In general we may also have a kinetic mixing term given
by $\kappa Z_{\mu \nu}{Z^{\prime}}^{\mu \nu}$. The value of
$\kappa$ is however severely constrained by electroweak
precision measurements ($\kappa \la 10^{-4}$ \cite{Hook:2010tw}). So
for calculational simplicity we have restricted ourselves
to a parameter space where $\kappa < \gbl$, hence neglecting
its contribution. These type of scenarios where kinetic mixing
term is neglected has been previously studied under the name
of ``Minimal/Pure''$\ubl$ model
\cite{1002.2525, 1308.0023, Basso:2008iv}.}
\begin{eqnarray}
\mathcal{L}_{\rm gauge} = \mathcal{L}^{\rm SM}_{\rm gauge}
-\dfrac{1}{4} Z^{\prime}_{\mu \nu} {Z^{\prime}}^{\mu \nu}\,.
\label{lagrang_gauge}
\end{eqnarray}
Here, $\mathcal{L}^{\rm SM}_{\rm gauge}$ is the Lagrangian of
the SM gauge sector while the second term represents the kinetic
term for the $\bl$ gauge bosons $\zbl$ and in terms of
$\zbl$ the field strength tensor ${Z^{\prime}}^{\mu \nu}$ for an abelian gauge
field is defined as
\begin{eqnarray}
{Z^{\prime}}^{\mu \nu} = \partial^{\mu} {\zbl}^{\nu}
- \partial^{\nu} {\zbl}^{\mu} \,.
\end{eqnarray}
The gauge invariant Lagrangian for the three right
handed neutrinos can be written as:
\begin{eqnarray}
\mathcal{L}_{\rm RN} =  i\,\sum_{i=1}^3\bar{N_i}{{D\!\!\!\!\slash}_N}
N_i - \lambda_{R_i}\bar{N^c_i}
N_{i} \Psi+\sum_{\alpha=1}^{3}\sum_{i=1}^{3}y_{\alpha i}
\bar{L_{\alpha}}\tilde{\Phi}N_{i}\,,
\label{lrn}
\end{eqnarray}
where $\tilde{\Phi}=-i\tau_{2}\Phi^{*}$ and
${{D\!\!\!\!\slash}_N}= \gamma_{\mu}\,D^{\mu}_N$ with
\begin{eqnarray}
D^{\mu}_N N_i &=& \left(\partial_{\mu} -
i\,{g_{\rm BL}}\,{Z_{\rm BL}}_{\mu} \right) N_i
\end{eqnarray}
is the covariant derivative for the right handed neutrino $N_i$.
After $\ubl$ symmetry breaking the masses of right handed neutrinos
and $\zbl$ are given by
\begin{eqnarray}
M_{Z_{\rm BL}}^{2}&=&4 g_{\rm BL}^2 v_{\rm BL}^2\,, \\
M_{N_{i}}&=&{\sqrt{2}}\lambda_{R_{i}}{v_{\rm BL}}\,.
\label{mzbl-mn}
\end{eqnarray}
Using above two equations one can write the coupling $\lambda_{R_i}$
in terms of $\gbl$, $M_{N_i}$ and $\mzbl$ which is 
\begin{eqnarray}
\lambda_{R_{i}} &=& \sqrt{2}\left(\frac{M_{N_{i}}}
{M_{Z_{\rm BL}}}\right)g_{\rm BL} \,.
\label{lambdar}
\end{eqnarray}
From Eq. (\ref{lrn}) it is possible to generate neutrino
masses via Type-I see-saw mechanism. In our analysis we
want to focus on the viability of lightest sterile neutrino ($N_1$)
as a dark matter candidate. So for simplicity we have neglected
intergenerational mixing between the active and sterile neutrinos.
The mass of the other two sterile neutrinos are also not constrained
by our analysis in this work, and in principle
can be very heavy aiding neutrino mass generation
by the see-saw mechanism. From Eq. (\ref{lrn}) and Eq. (\ref{mzbl-mn}),
one can find the expression of active-sterile mixing angle $\alpha_i$ per
generation as
\begin{eqnarray}
\tan 2\,\alpha_i = -\dfrac{\sqrt{2}\,y_{i}\,v}{M_{N_i}}\,.
\end{eqnarray} 
For simplicity, throughout this work we have denoted the first generation
active-sterile mixing angle $\alpha_1$ by only $\alpha$.

The non-observation of the extra neutral gauge boson in the LEP experiment
\cite{hep-ph/0408098,hep-ph/0604111} imposes following constraint
\footnote{For recent bounds on $\mzbl$ and $\gbl$ from the LHC
experiment see Ref. \cite{Aad:2014cka}.}
on the ratio of $\mzbl$ and $\gbl$:
\begin{eqnarray}
\frac{M_{Z_{\rm BL}}}{g_{\rm BL}}=2v_{\rm BL} \geq 6-7\,\, {\rm TeV}.
\end{eqnarray}
In our analysis independent parameters are:\\
Mass of the extra singlet Higgs $M_{H}$, Masses of all
three RH neutrinos $M_{N_{i}}$, mass of extra neutral gauge boson
$M_{Z_{\rm BL}}$, scalar mixing angle $\theta$, the new gauge
coupling $g_{\rm BL}$ and active-sterile mixing angle $\alpha$.
In terms of our chosen independent set of model parameters, the other
parameters appearing in Eq. (\ref{scalar-potential}) can be written as
\begin{eqnarray}
\mu_{1}^2&=&-\frac{v \left(M_{h}^2+M_{H}^2\right)+(M_{h}^2-M_{H}^2)
(v \cos 2\, \theta -v_{BL} \sin 2\, \theta )}{4\,v}\,,\\
\mu_{2}^2&=&\frac{-v^3 \left(M_{h}^2+M_{H}^2\right)+(M_{h}^2-M_{H}^2)
\left(v^3 \cos 2\, \theta +v_{\rm BL}^3 \sin 2\,\theta \right)}
{4 v\,v_{\rm BL}^2}\,,\\
\lambda_{1}&=&\frac{M_{h}^2+\cos 2\, \theta  (M_{h}^2-M_{H}^2)+M_{H}^2}{4\,v^2}\,,
\label{lam1}\\
\lambda_{2}&=&\frac{\cos 2\, \theta  \left(M_{H}^2-M_{h}^2\right)+
M_{h}^2+M_{H}^2}{4\,v_{\rm BL}^2}\,,
\label{lam2}\\
\lambda_{3}&=&\frac{\sin \theta  \cos \theta 
\left(M_{H}^2-M_{h}^2\right)}{v\,v_{\rm BL}}\,.
\label{lam3}   
\end{eqnarray}
\section{Exploring the Non-thermal Regime}\label{NTR}
Non-thermal production mechanism of dark matter has been studied
for quite a long time. Their characteristic behaviour comes
from the very low cross section with the Standard Model particles in
the early Universe. Due to this very low cross section
(lower than that of WIMPs), the non-thermal dark matter particles
can never reach in thermal equilibrium with the Standard Model particles.
Hence their evolution in the early Universe is studied differently
than the thermal scenario. In the thermal scenario, the abundance of
a relic particle (called WIMP) remains nonzero in the present epoch 
due to the ``Freeze-out'' mechanism \cite{Gondolo:1990dk}, whereas
in the case of a non-thermal production of DM (called FIMP),
a different mechanism known as ``Freeze-in'' \cite{0911.1120}
is responsible for their relic abundance.
In the non-thermal case, due to very low interaction cross section, the initial
abundance of the dark matter is taken to be zero. As the Universe
cools, they are dominantly produced by the decay of other SM/BSM
particles. They can also be produced by the scattering of SM/BSM
particles, but with a sub-dominant contribution. Once the non-thermal
dark matter is produced, due to extremely low interaction strength,
they do not thermalise with the rest of the thermal soup.
Since most of the production of DM particles in the non-thermal
regime occur from the decays of heavier particles, non-thermality
condition will be satisfied when the rate
of production from the decaying mother particle (decay width)
is less than the expansion rate of the Universe at around a temperature
$T\sim M$, where $M$ is the mass of the decaying particle \cite{Arcadi:2013aba}.  
Mathematically this can be written as
\begin{equation}
~~~~~~~~~~~~~~~~~~\frac{\Gamma}{H} < 1\,\,\,\,\,\,(\text{for}\,\,T\sim M)\,,
\label{nt}
\end{equation}
where, $\Gamma$ is the relevant decay width and $H$ is the
Hubble parameter. However in some cases, if the production
of DM particles may occur mainly from the annihilation of
other particles in the thermal bath (production from decay
can be forbidden due to kinematical condition or by some symmetry
in the Lagrangian). $\Gamma$ will then be replaced by:
\begin{equation}
\Gamma = n_{eq} \langle {\sigma {\rm v}} \rangle\,,
\end{equation}
where, $\langle {\sigma {\rm v}} \rangle$ is
the thermally averaged annihilation cross section
of the particles in the thermal bath and 
$n_{eq}$ is their {\it equilibrium} number density.

In this U(1)$_{\rm B-L}$ model, to calculate the relic density
of a non-thermal sterile neutrino dark matter ($N_1$),
the principal ingredient is its production from
various decay and annihilation channels. This gives the required
comoving number density of $N_1$ upon solving the relevant Boltzmann equation.
The main production channels of the sterile neutrino (via decay) are :\\
$W^{\pm} \rightarrow N_{1}\,e^\pm$, $Z\rightarrow N_{1} \bar{N_1}$,
$Z_{\rm BL}\rightarrow N_1 \, \bar{N_1}$, $H \rightarrow N_1 \bar{N_1}$,
$ h \rightarrow N_1 \bar{N_1}$.\\
The corresponding decay widths are given in the Appendix \ref{decay_channel}.
As discussed earlier, non-thermal dark matter particles
can also be produced from the
scattering of the SM/BSM particles in the thermal soup. The rate of
the back reactions are negligible, since the number density of
$N_1$ is extremely small in the early Universe. The annihilation
channels along with corresponding cross sections aiding the
production of $N_1$ are also given in the Appendix \ref{annihilation_channel}.
As we will see later, in the present case $W^\pm$ and $\zbl$ decays are
main production channels of $N_1$. Using the non-thermality
condition given in Eq. (\ref{nt}) we find that the extra
gauge coupling $\gbl$ and the active-sterile
mixing angle $\alpha$ must be less than $10^{-9}$
and $10^{-7}$ (rad) respectively for an $\mathcal{O}$(MeV)
sterile neutrino with the mass
of $\zbl$ lying in 1 GeV to 100 GeV range.
Although this is a simple way to estimate the order
of magnitude of $\gbl$ and $\alpha$
required for the dark matter candidate ($N_1$)
to be non-thermal, this sets a very first upper
limit on these quantities. However, more stringent upper bound on
$\alpha$ ($\alpha\la10^{-9}$ rad) arises from the stability of
DM over the cosmological time scale. 

Moreover, an upper bound on the active-sterile mixing angle
$\alpha$ is also obtained from the invisible decay of
the Standard Model $Z$ boson. Following Ref.\,\cite{Agashe:2014kda} we find:
\begin{eqnarray}
\dfrac{\Gamma(Z \rightarrow \text{inv})}{\Gamma(Z \rightarrow \nu \nu)}
\,=\,2.990 \pm 0.007\,.
\end{eqnarray}
In the limit when active-sterile mixing angle is
small and $M_{Z} \gg \mdm$, from the above equation
we get $\sin ^{4} \alpha < 0.007$. As we will see later
that for us, this condition is indeed being satisfied.
In the present scenario since $M_h < 2 \, M_H$,
SM Higgs boson can decay {\it invisibly} only into a pair of
lightest sterile neutrino $N_1$. From the expression of
the decay width given in Eq. (\ref{hn1n1}) we find that
it is suppressed by $\gbl^2$ and hence very small.
Thus this decay width easily satisfies the bound on
invisible decay of SM Higgs boson from
LHC \cite{Bechtle:2014ewa}.
Furthermore, due to sufficiently
small interaction strength with the SM particles,
non-thermally produced $N_1$ always satisfies all
the existing bounds on spin independent as well as spin dependent
scattering cross sections from dark matter direct
detection experiments \cite{1512.03506}.
 
We have mentioned earlier that for the non-thermal production of the
sterile neutrinos, the coupling constant $\gbl$ should be very small
($<$10$^{-9}$). As is usually done, while considering the
production of dark matter from a decay of any SM/BSM particle, the
latter is implicitly assumed to be in thermal equilibrium. Hence we
usually do not need to solve a system of {\it coupled} Boltzmann
equations, since the equilibrium number density is assumed for the
decaying mother particle. But, here due to very low interaction
strength of $Z_{\rm BL}$ (due to small $\gbl$), it will not
be in thermal equilibrium with the rest of the
particles. Also, the decay of $\zbl$ is a mode of
production of the our sterile neutrino dark matter $N_1$.
So, first, we find the comoving number density of $\zbl$
by solving its Boltzmann equation. Then we use this to find the
relic density of our sterile neutrino dark matter.
Thus, in our case we have to solve a set of two {\it{coupled}}
Boltzmann equations, one for the sterile neutrino dark matter,
and another for the $Z_{\rm BL}$.  

In any model with a sterile neutrino we will have an active-sterile
mixing in general. Hence in such model {\it{production}} of the sterile
neutrino via $W^\pm$ decay is a very generic feature. But, it is usually
not taken into account since it is suppressed by the square of
the small active-sterile mixing angle. However, in this work, in
our favoured parameter space, we find that a sizeable contribution
(to the relic-density of $N_1$) even from the $W^\pm$ decay
is present (see Section \ref{sec_BE_full}).\\
Another important feature which will be present for a generic model
having an nonzero active sterile mixing is the production of sterile
neutrino through the Dodelson-Widrow (DW) mechanism. Here the production
of sterile neutrino occurs via the oscillations of active neutrinos to
the sterile ones. But this mechanism suffers serious drawbacks
from the Lyman-$\alpha$ bounds \cite{0812.0010} as well as X-ray
observations \cite{1309.4091}. It is now known
\cite{astro-ph/0602430, 0812.3256} that sterile neutrino
produced by this mechanism cannot comprise the whole of the dark matter
of the Universe. The contribution arising to the relic abundance of
a sterile neutrino from the DW mechanism is given by \cite{astro-ph/0101524}
\begin{eqnarray}
\Omega_{\rm DW}h^2 \approx 0.3\times\left(\frac{\sin^2 2\,\alpha}
{10^{-10}}\right)\left(\frac{\mdm}{100\,\rm keV}\right)^2\,,
\label{DW}
\end{eqnarray}
where, $\alpha$ is the active sterile mixing angle and
$\mdm$ is the mass of the sterile neutrino. In our case we find
(see Section \ref{sec_RD} for details) that in order to satisfy
relic density, $\alpha$ should be less than $10^{-10}$ rad for 
sterile neutrino mass lying between 1 MeV and 10 MeV.
From Eq. (\ref{DW}) we see that the corresponding DW
contribution to the relic density is $\la 1.2 \times 10^{-6}$ and
hence negligible. 
\section{Boltzmann Equation}
\label{sec_BE}
In this section, we write the two coupled Boltzmann
equations that dictates the final relic abundance of the sterile neutrino
dark matter $N_1$. The Boltzmann equation for the evolution of $Z_{\rm BL}$
which, as already discussed is very weakly interacting is given by
\footnote{In general the first term of Eq. (\ref{bltz1}) will
look like: $\langle \Gamma_{H\rightarrow Z_{\rm BL}Z_{\rm BL}}
\rangle (Y_H^{eq} - Y_{Z_{\rm BL}})$, but since the initial abundance of
$Z_{\rm BL}$ is very small, we have neglected the inverse process
i.e. $Z_{\rm BL}\,Z_{\rm BL} \rightarrow H$, and consequently
dropping the $\langle \Gamma_{H\rightarrow Z_{\rm BL}Z_{\rm BL}}
\rangle Y_{Z_{\rm BL}}$ term in our analysis.}:
\begin{eqnarray}
\frac{dY_{Z_{\rm BL}}}{dz} &=&\frac{2M_{pl}}{1.66\,M_h^2}
\frac{z \sqrt{g_\star(z)}}{g_s(z)}
\Bigg(\langle \Gamma_{H\rightarrow Z_{\rm BL}Z_{\rm BL}} \rangle Y_H^{eq} -
\langle \Gamma_{Z_{\rm BL} \rightarrow all} \rangle Y_{Z_{\rm BL}}\Bigg)
\,.\nonumber \\
\label{bltz1}
\end{eqnarray}
Here, $Y_{Z_{\rm BL}} \equiv \dfrac{n_{Z_{\rm BL}}}{\rm s}$ is the comoving
number density of the extra gauge boson with $n_{Z_{\rm BL}}$ and $\rm s$
being the number density of $Z_{\rm BL}$ and the entropy density of the Universe
respectively. Also $z \equiv \dfrac{\Lambda}{T}$ where $\Lambda$ is a mass scale
and $T$ is the temperature of the Universe. For simplicity we have taken
$\Lambda \sim M_h$, the mass of SM Higgs boson while $M_{pl}$ is the usual
Planck mass. The function $g_\star(z)$ is given by:
\begin{eqnarray*}
\sqrt{g_\star(z)} = \frac{g_{\rm s}(z)}{\sqrt{g_{\rho}(z)}}
\left(1 - \frac{1}{3}\frac{d\,{\rm ln}\,g_{\rm s}(z)}{d\,{\rm ln}z}\right) \, ,
\end{eqnarray*}
where, $g_{\rho}(z)$ and $g_{\rm s}(z)$ are the effective degrees of freedom
related to the energy density $\rho$ and the entropy density $\rm s$ of the Universe
respectively. The quantity $\langle \Gamma_{A \rightarrow BB} \rangle$
denotes the thermally averaged decay width for the process $A \rightarrow BB$
and its expression, in terms of decay width $\Gamma_{A \rightarrow BB}$, is
given by \cite{1404.2220} \footnote{For a more rigorous approach,
when the decaying particle is $\zbl$, one should use its non-thermal
distribution function ($f_{\zbl}$) for calculating this thermally
averaged decay width. The expression will look like:
$\langle \Gamma_{\zbl \rightarrow \, BB} \rangle = \dfrac{\int\,
(\frac{M_{\zbl}}{E_{\zbl}})\,\Gamma_{\zbl \rightarrow BB}
f_{\zbl}(p,T)\,d^3p}{\int\,f_{\zbl}(p,T)\,d^3p}$.
The non-thermal distribution function $f_{\zbl}$
should be obtained first by solving the appropriate Boltzmann equation.}:

\begin{eqnarray}
\langle \Gamma_{A \rightarrow BB}\rangle &=&
\dfrac{K_1(z)}{K_2(z)}\Gamma_{A\rightarrow BB}\,.
\end{eqnarray}
Here, $K_1(z)$ and $K_2(z)$ are the modified Bessel functions of order 1 and 2
respectively. The expressions for the relevant decay widths are given in Appendix
\ref{total_decay}.

\begin{figure}[h!]
\centering
\includegraphics[scale=0.5,angle=-90]{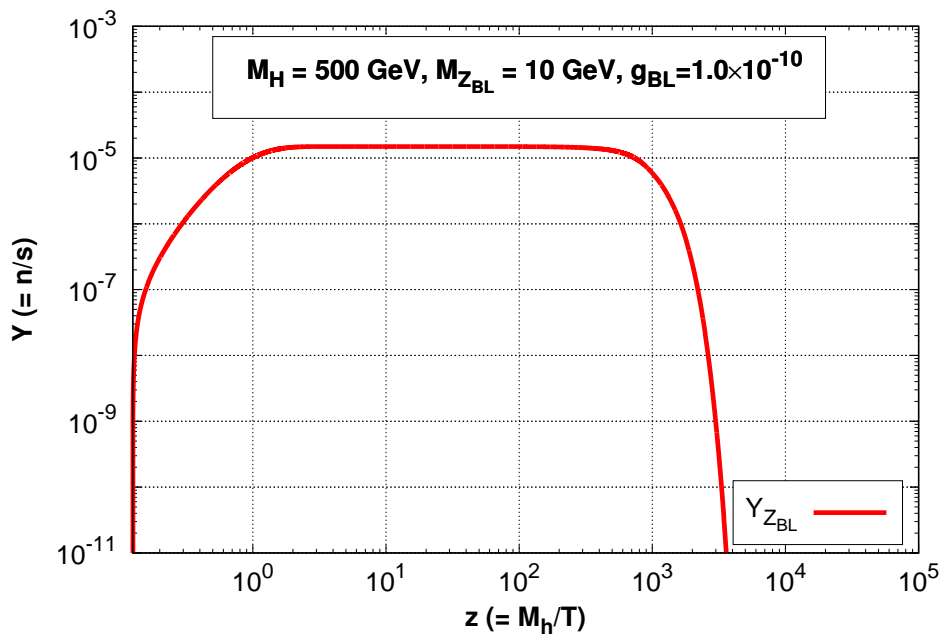}
\caption{Evolution of comoving number density of $Z_{\rm BL}$ with respect to $z$.}
\label{zbleqn}
\end{figure}

The SM particles acquire their masses after the process of EWSB 
whereas the BSM particles like U(1)$_{\rm B-L}$ gauge boson $Z_{\rm BL}$
and the additional Higgs boson ($H$) gain their masses after
the breaking of U(1)$_{\rm B-L}$ symmetry. Therefore, in the early
Universe the main production channel of the new gauge boson
is mainly through the decay of $H$, while the latter is 
in thermal equilibrium with the plasma. 
The first term in Eq. (\ref{bltz1})
denotes this contribution to the production of $Z_{\rm BL}$
(i.e. increase in number density of $Z_{\rm BL}$)
and hence comes with a positive sign.
Since in our case,
both the masses of $Z_{\rm BL}$ and $H$ are free parameters,
we have adopted the values of $M_H$ in a range such that it
always satisfy the kinematical condition $M_H \geq 2\, M_{Z_{\rm BL}}$.
Although in the early stage of the Universe, the decay of $H$ is the main
production channel of $Z_{\rm BL}$, in principle it can also be produced
from the annihilation processes, involving both SM as well as BSM
particles, like $hh \rightarrow Z_{\rm BL}Z_{\rm BL}$,
$W^+W^- \rightarrow Z_{\rm BL}Z_{\rm BL}$,
$ZZ \rightarrow Z_{\rm BL}Z_{\rm BL}$,
$HH \rightarrow Z_{\rm BL}Z_{\rm BL}$,
$N_{2,\,3} \bar{N}_{2,\,3} \rightarrow Z_{\rm BL}Z_{\rm BL}$ etc.
However, contribution of these annihilation processes
is subleading to that of decay.
The number density of extra gauge boson $Z_{\rm BL}$ is also
depleted mainly through its decay modes to $N_1 \bar{N_1}$
(other two sterile neutrinos are assumed to be heavy for simplicity),
$\nu_x \bar{\nu_x}$ and $f \bar{f}$. It is denoted by the
second term in the Boltzmann equation (Eq. (\ref{bltz1})),
and as expected it comes with a negative sign,
since it signifies the depletion of $Z_{\rm BL}$ number density.
These two competing processes (production vs. depletion)
decide the final comoving number density of $Z_{\rm BL}$.
Numerically solving Eq. (\ref{bltz1}), we graphically show
the evolution of comoving number density of $Z_{\rm BL}$ with
$z=\frac{M_h}{T}$ in Fig. \ref{zbleqn}.
\\
From the above plot it is seen that, the comoving number density
of $\zbl$ first rises due to the production term (first term in the R.H.S. of
Eq. (\ref{bltz1})) and then after a certain time it falls
when the depletion term (i.e. the second term in the R.H.S. of Eq. (\ref{bltz1}))
begins to dominate. This situation arises because, at that time the temperature
of the Universe becomes much smaller than $M_H$ ($T\ll M_H$) and hence,
being a non relativistic species, the equilibrium number density of
$H$ is exponentially suppressed. Therefore, the production of
$\zbl$ ceases. The middle ``plateau'' like portion occurs
when both the production and the depletion terms are
comparable and hence compensating each other.
The plot is generated for the following chosen set of relevant parameters: 
$M_{H}=500$ GeV, $M_{Z_{\rm BL}}=10$ GeV and $g_{\rm BL}=10^{-10}$.

Now, we proceed to write the Boltzmann equation of the lightest sterile neutrino
$N_1$. This will govern the number density of the dark matter candidate ($N_1$)
and consequently its relic abundance at the present epoch.
Similar to Eq. (\ref{bltz1}), the Boltzmann equation for $N_1$
is given by :
\begin{eqnarray}
\frac{dY_{N_1}}{dz} &=&\frac{2M_{pl}}{1.66\,M_h^2}
\frac{z \sqrt{g_\star(z)}}{g_s(z)}
\Bigg(\langle \Gamma_{W^\pm \rightarrow e^\pm N_1} \rangle(Y_W^{eq} -Y_{N_1})
+\langle \Gamma_{Z_{\rm BL} \rightarrow N_1 N_1}
\rangle(Y_{Z_{\rm BL}}-Y_{N_1})\nonumber\\
&& + \sum_{i=H,h,Z} \langle \Gamma_{i \rightarrow N_1 N_1}\rangle
(Y_{i}^{eq}-Y_{N_1})\Bigg)+\frac{4 \pi^2}{45} 
\frac{M_{pl} M_{h}}{1.66}\frac{\sqrt{g_{\star}(T)}}{z^2} \times
\nonumber \\ &&
\Bigg(\sum_{x = W, Z, f, H}
\langle {\sigma {\rm v}}_{x\bar{x}\rightarrow N_1 N_1}\rangle
\,\,{({{Y}_x^{eq}}\,^2 -Y_{N_1}^2)}+
\langle {\sigma {\rm v}}_{Z_{\rm BL}Z_{\rm BL}\rightarrow N_1 N_1}\rangle
\,\,({{Y}_{Z_{\rm BL}}^2 -Y_{N_1}^2)}
\Bigg)\,\,.\nonumber \\
\label{bltz2}
\end{eqnarray}
As discussed in Eq. (\ref{bltz1}), since the initial abundance
of the sterile neutrino dark matter $N_1$ is very small, the $Y_{N_1}$
term in the above equation may be neglected \cite{0911.1120, 1306.3996}.
Here $\langle {\sigma {\rm v}}_{x\bar{x}\rightarrow N_1 N_1}\rangle$ is the
thermally averaged cross section for the production of
$N_1$ from the annihilation of $x$ particle. The expression of
$\langle {\sigma {\rm v}}_{x\bar{x}\rightarrow N_1 N_1}\rangle$
is given by \cite{Gondolo:1990dk} \footnote{As previously discussed,
in a strict sense, one should use a definition of
$\langle {\sigma {\rm v}}_{Z_{\rm BL}Z_{\rm BL}\rightarrow N_1 N_1}\rangle$
based on the non-equilibrium density function $f_{\zbl}$.}
\begin{eqnarray}
\langle {\sigma {\rm v}_{x\bar{x}\rightarrow N_1 N_1}} \rangle =
\frac{1}{8 M_x^4 T K_2^2\left(\frac{M_x}{T}\right)}
\int_{4M_x^2}^\infty \,\sigma_{x x\rightarrow N_1 N_1}\,
(s-4M_x^2)\,\sqrt{s}\,K_1\left(\frac{\sqrt{s}}{T}\right)\,ds \,\,.
\nonumber\\
\end{eqnarray}
The expressions for the relevant decay widths and annihilation cross sections
are given in the Appendix \ref{decay_channel} and \ref{annihilation_channel}
respectively.

\begin{figure}[h!]
\centering
\includegraphics[scale=0.5,angle=-90]{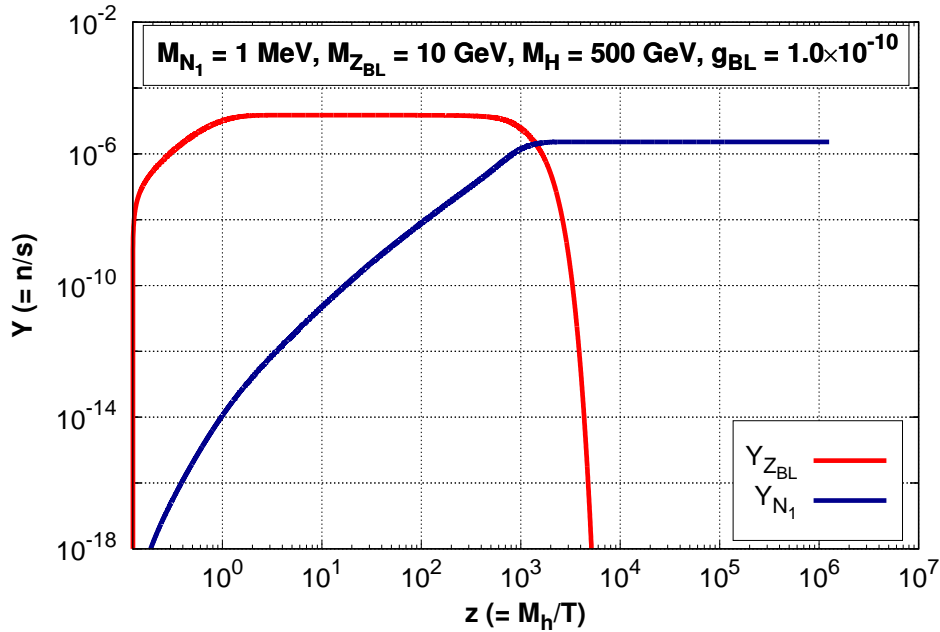}
\caption{Evolution of comoving number densities of $\zbl$ and $N_1$.}
\label{plot2}
\end{figure}

In order to get the comoving number density ($Y_{N_1}$) of $N_1$
at the present epoch, we have to solve the coupled set of
Boltzmann equations given in Eq. (\ref{bltz1}, \ref{bltz2}).
To be more precise, the value of $Y_{Z_{\rm BL}}(z)$ for each $z$
obtained by solving Eq. (\ref{bltz1}) is to be fed into
Eq. (\ref{bltz2}). For understanding the physics
behind this coupled set of Boltzmann equations better,
let us first assume that the lightest sterile neutrinos
are {\it{only}} produced from the decay of $Z_{\rm BL}$,
whereas $Z_{\rm BL}$ is produced and depleted according to Eq. (\ref{bltz1}).
We plot the result in Fig. \ref{plot2}.
In Fig. \ref{plot2}, initially at the early stage of
the Universe there are no $Z_{\rm BL}$ particles
and hence no $N_1$, since for simplicity we have switched off all
other production channels of $N_1$, except $Z_{\rm BL}$. Then,
when $Z_{\rm BL}$ is produced from the decay of $H$,
we also find an increase in the number density of $N_1$
from the decay of $Z_{\rm BL}$. Finally, the number
density of $Z_{\rm BL}$ begins to fall due to its dominating
decay modes (production of $\zbl$ terminates as the the number
density of $H$ becomes negligibly small), and consequently
the number density of $N_1$ also saturates since now there
are no $Z_{\rm BL}$ left to aid the production of $N_1$.
This plot is also generated for the following chosen set
of relevant parameters: $M_{H}=500$ GeV, $M_{Z_{\rm BL}}=10$ GeV,
$M_{N_1}=1$ MeV and $g_{\rm BL}=10^{-10}$.

\begin{figure}[h!]
\centering
\includegraphics[scale=0.5,angle=-90]{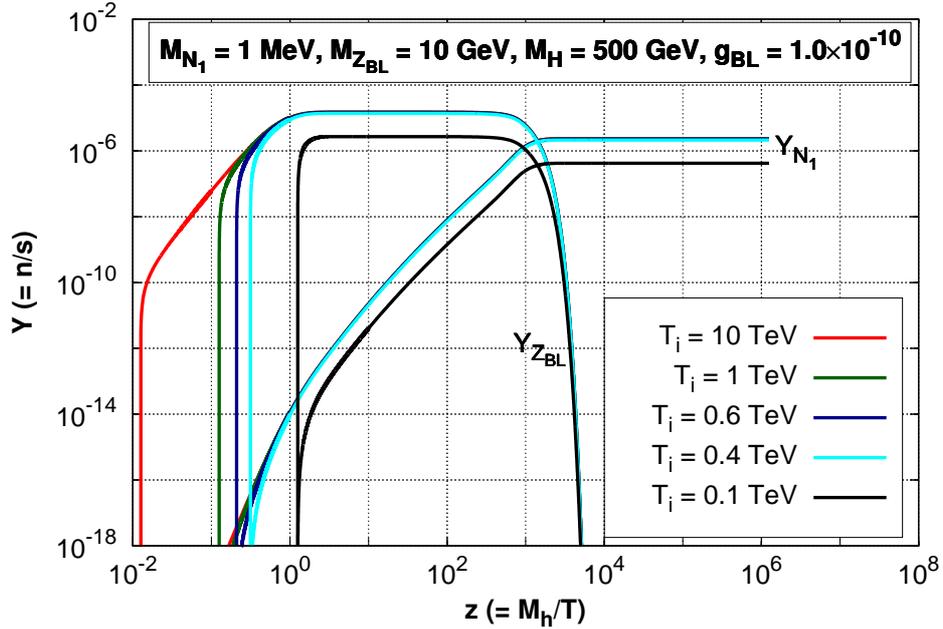}
\caption{Dependence of $Y_{\zbl}$ and $Y_{N_1}$ on different
sets of initial temperatures.}
\label{initemp}
\end{figure}

In Fig. \ref{plot2} we have taken the initial temperature ($T_i$)
to be 1 TeV. The final abundances of $\zbl$ and $N_1$ will not depend on
this as long as $T_i \ga M_H$. The maximum production of $\zbl$ from
$H$ decay occurs around a temperature of $\sim M_H$. However, if $T_i$
becomes less than $M_H$ then, since $H$ is in thermal equilibrium,
its own abundance will be exponentially suppressed (as it becomes
non-relativistic) and thereby reducing $Y_{\zbl}$ and $Y_{N_1}$.
All these are shown in Fig. \ref{initemp} where we see as discussed above,
for $T_i \ga M_H$, there is no change in the final values of
$Y_{\zbl}$ and $Y_{N_1}$ (red, green, blue, cyan solid lines).
While for $T_i \la M_H$ both the final abundances are
reduced (black solid line) from their previous values.

We now show the variation of Fig. \ref{plot2}
with different sets of chosen model parameters.
\begin{figure}[h!]
\centering
\subfigure[Variation with different $\gbl$ values]
{\includegraphics[scale=0.3,angle=-90]{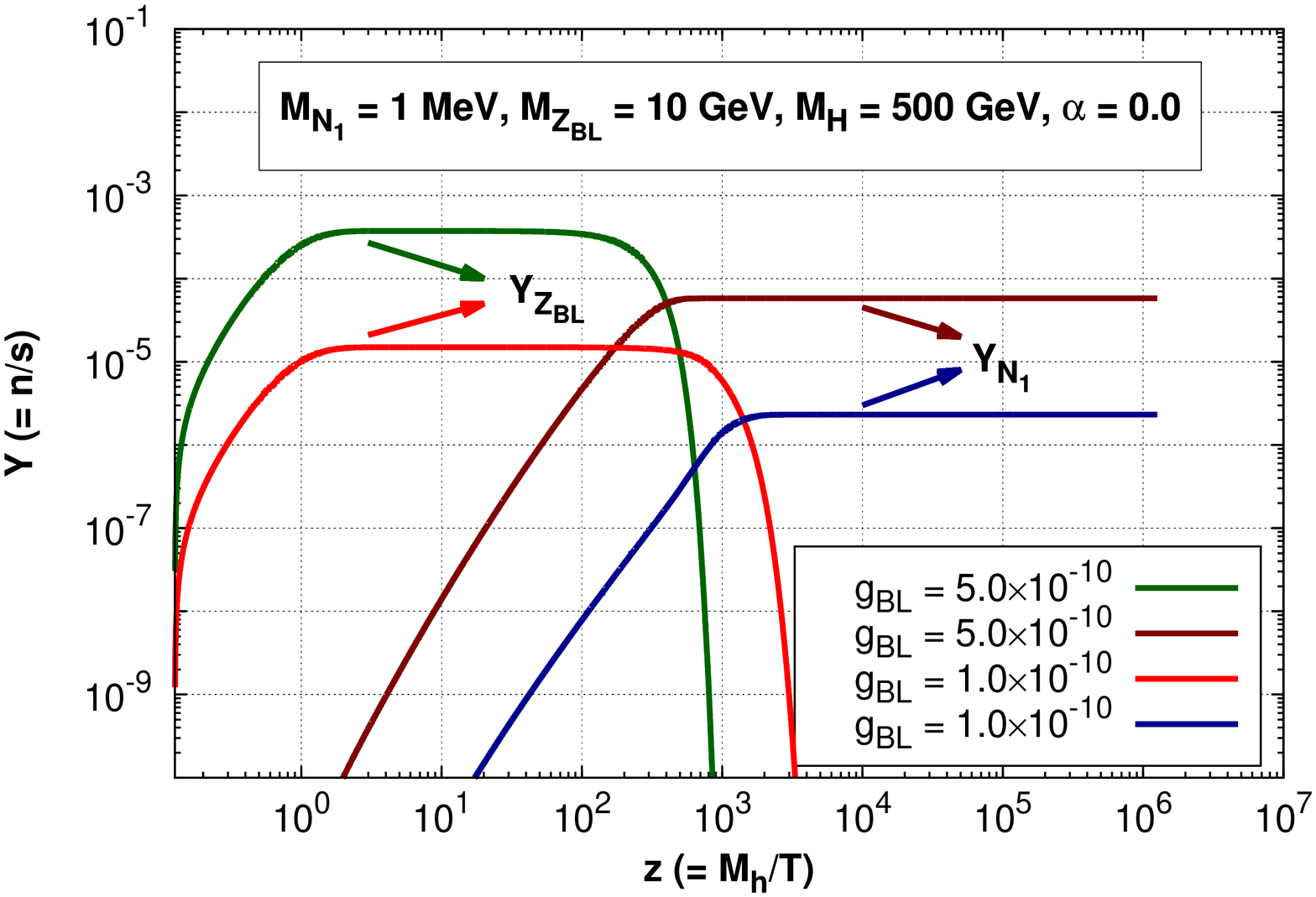}}
\subfigure[Variation with different $M_H$ values]
{\includegraphics[scale=0.3,angle=-90]{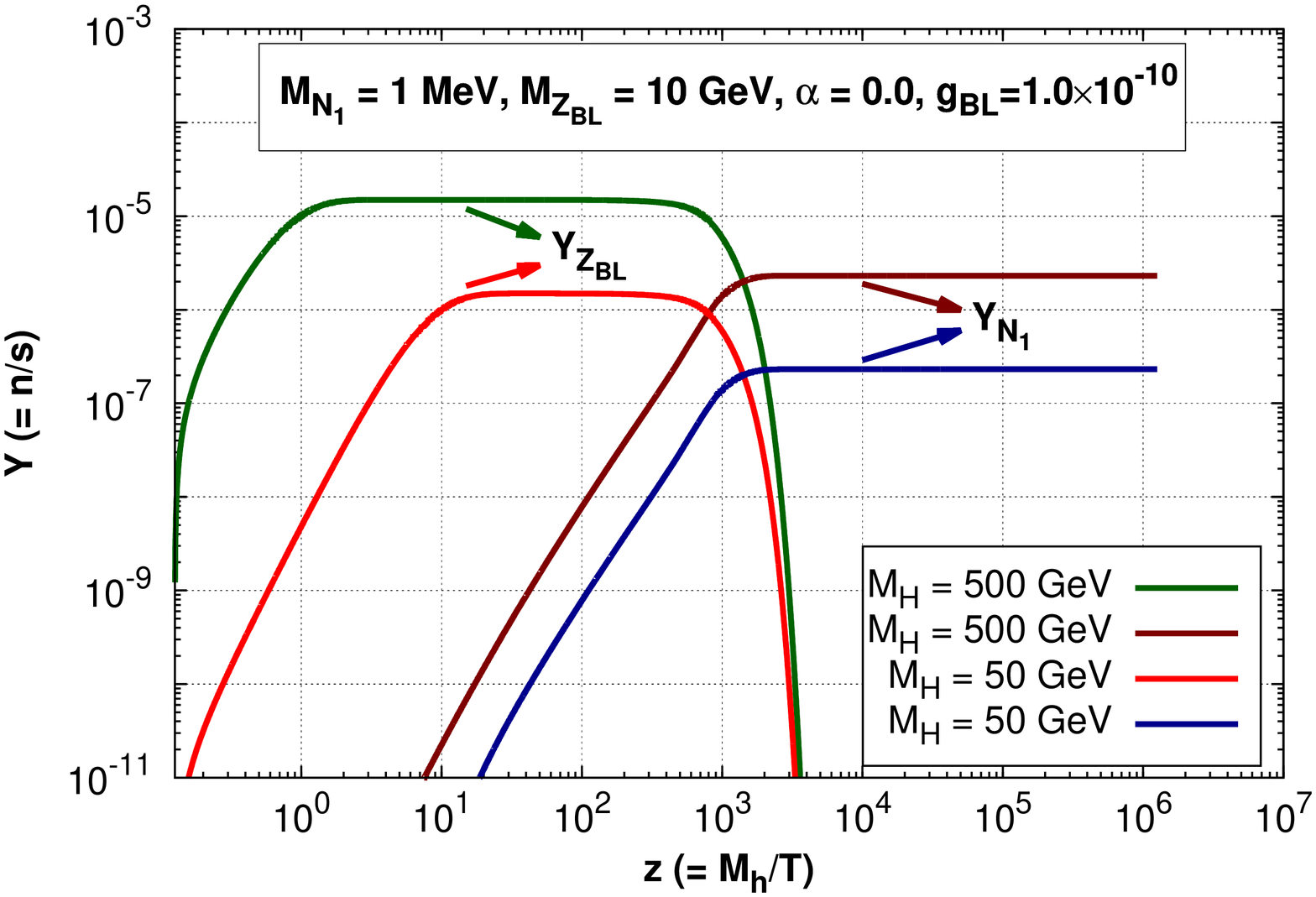}}\\
\subfigure[Variation with different $\mdm$ values]
{\includegraphics[scale=0.3,angle=-90]{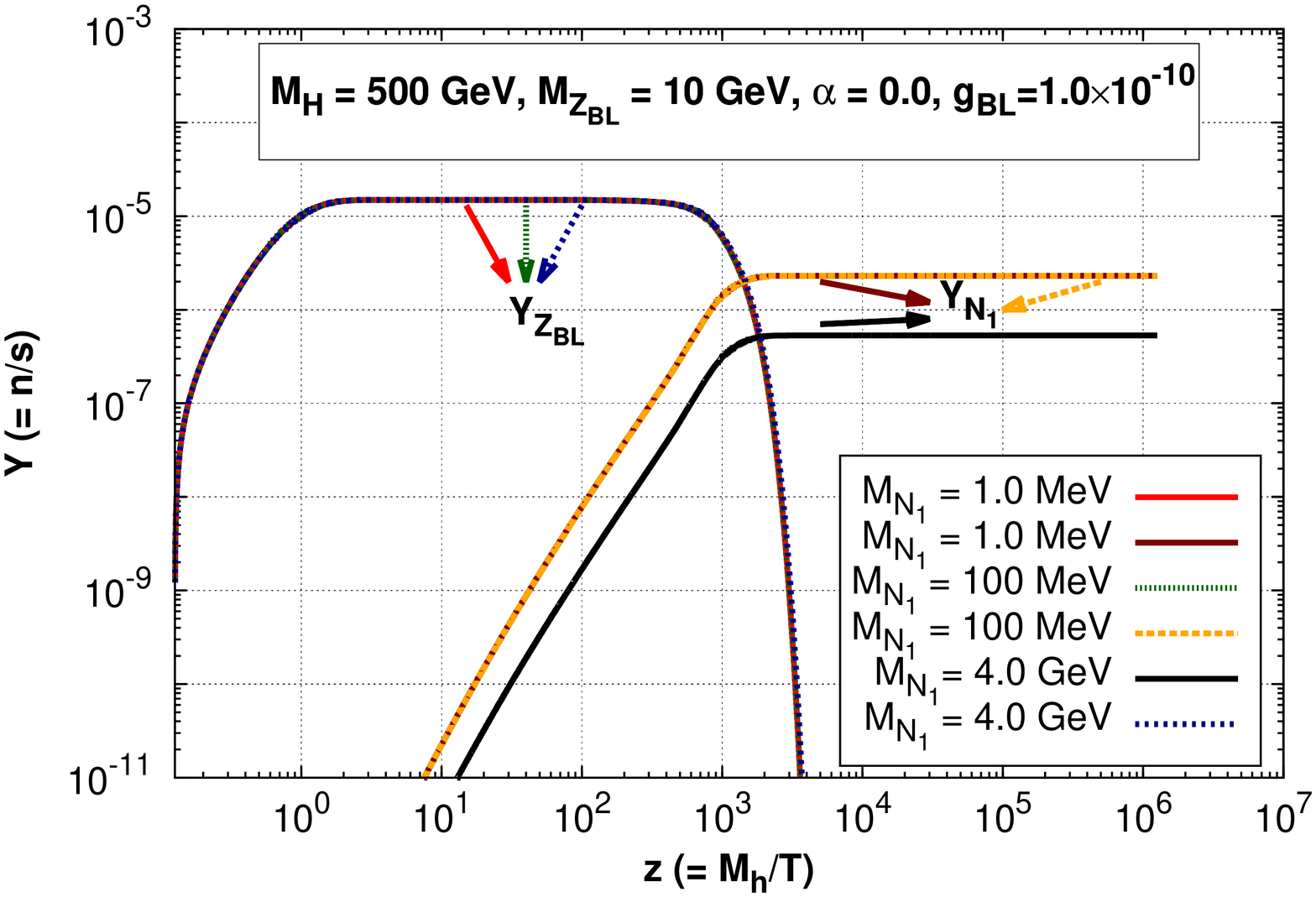}}
\subfigure[Variation with different $\mzbl$ values]
{\includegraphics[scale=0.3,angle=-90]{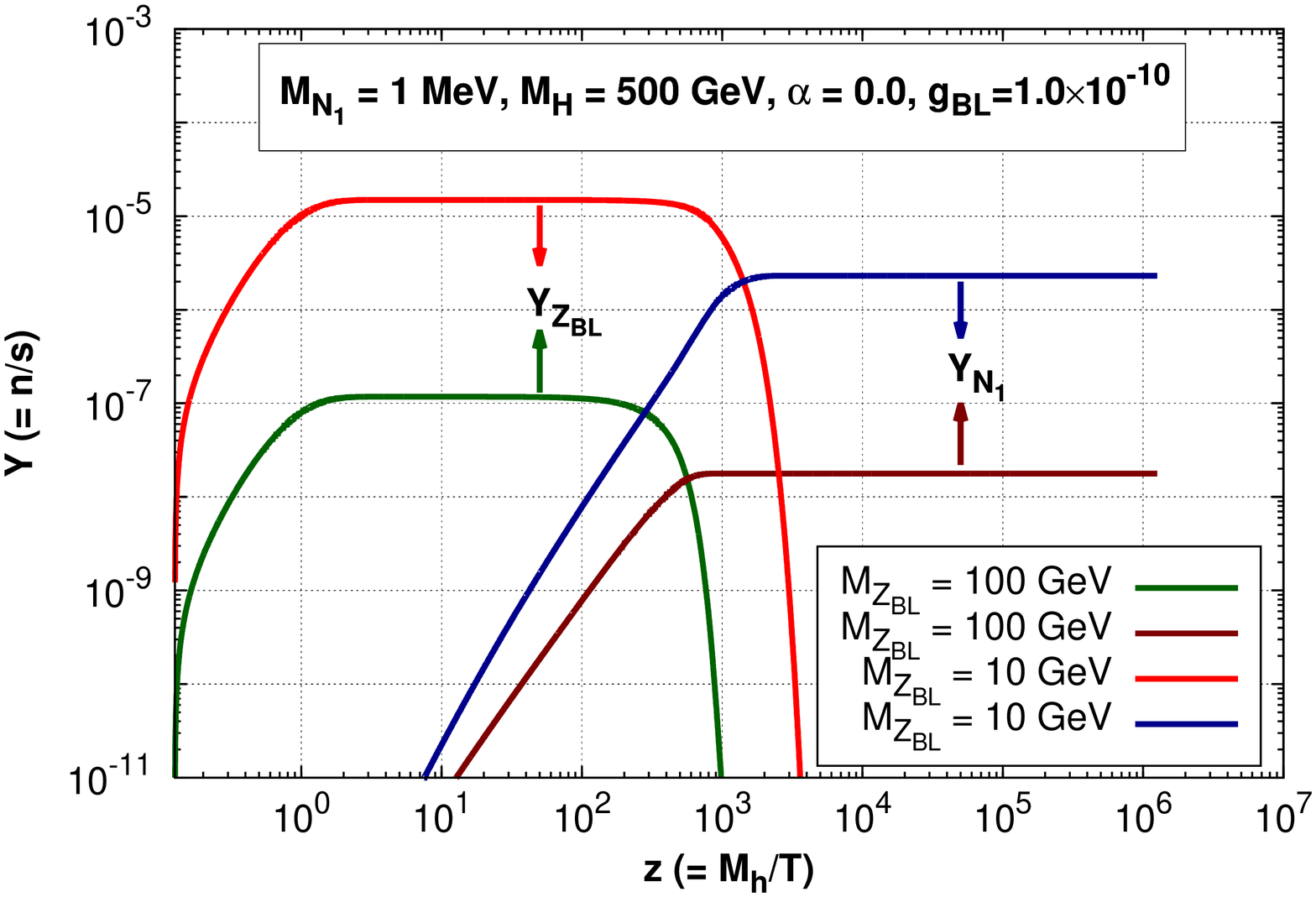}}
\caption{Comparison of comoving number densities of $Z_{\rm BL}$ and $N_1$
with respect to different sets of chosen model parameters.}
\label{plot3}
\end{figure}
In Fig.\,\ref{plot3}(a) we show the variation of $Y_{\zbl}$ and $Y_{N_1}$ with
$z$ for two different values of U(1)$_{\rm B-L}$ gauge coupling $\gbl$.
In this plot, we find that with increase in the value of $g_{\rm BL}$
the number density of $Z_{\rm BL}$ also increases initially.
This is understandable, because the decay width
$\Gamma_{H \rightarrow Z_{\rm BL}Z_{\rm BL}}$ increases with
$g_{\rm BL}$, and hence resulting in an increased number density
of the extra gauge boson. But the depletion rate of $Z_{\rm BL}$
(proportional to its total decay width) also increases with $g_{\rm BL}$,
and hence will result in a faster fall of its comoving number density.
This is also evident from the figure where the green line starts
to fall earlier than the red one. On the other hand, the production
rate of sterile neutrino dark matter ($N_1$) is proportional
to $\Gamma_{\zbl \rightarrow N_1 N_1}$ and $Y_{\zbl}$ (see Eq. (\ref{bltz2}))
and both of these quantities increase with $\gbl$. Therefore,
the comoving number density of $N_1$ increases as
the value of $\gbl$ changes from $1\times 10^{-10}$ to $5\times10^{-10}$.\\
In Fig. \ref{plot3}(b), we show the variation of $Y_{\zbl}$ and $Y_{N_1}$
with $z$ for two different values of $M_H$. Now, with a decrease in
$M_H$, we expect a corresponding decrease in decay width
$\Gamma_{H \rightarrow Z_{\rm BL}Z_{\rm BL}}$ ($\zbl$ production rate),
and hence the initial number density of $Z_{\rm BL}$ will be smaller.
This feature is seen in plot (b) where initially ($z \la 10^3$)
$Y_{\zbl}$ for $M_H=500$ GeV (green line) is larger than
that for $M_H=50$ GeV (red line). However, the total
decay width of $\zbl$ (and consequently its depletion rate)
does not depend on the mass of $H$. Hence both the red and green
lines will start to fall off around the same time.
Another noticeable change due to the variation of
$M_H$ is that the width of the ``plateau'' becomes
narrower with the decrease in mass difference
between $M_H$ and $M_{\zbl}$. 
Further, as the $Y_{\zbl}$ increases with an increase in
$M_H$, which in turn produces more $N_1$ (from the decay of $\zbl$)
in the final state and hence the comoving number density
$Y_{N_1}$ also increases with $M_H$.
\\
In Fig. \ref{plot3}(c) we have shown the variation of
$Y_{\zbl}$ and $Y_{N_1}$ for three different values of
sterile neutrino dark matter mass. From this plot we find
that there is not much variation in $Y_{Z_{\rm BL}}$
with changing $\mdm$. This is because the decay width
$\Gamma_{Z_{\rm BL} \rightarrow N_1 N_1}$ is subdominant 
with respect to the other decays modes of $\zbl$. Increasing
the mass of $N_1$ will lead to a further decrease of
$\Gamma_{Z_{\rm BL} \rightarrow N_1 N_1}$ and hence
will not affect the depletion rate  which
is dominantly controlled by the other decay
channels of $\zbl$. But since in this case (when other
production channels of $N_1$ are switched off)
$\Gamma_{Z_{\rm BL} \rightarrow N_1 N_1}$ solely controls the
production rate of $N_1$, $Y_{N_1}$ changes with $\mdm$.  
It is seen from Fig. \ref{plot3}(c) that if we increase the
sterile neutrino mass from 1 MeV to 4 GeV ($\mdm$ tends to $M_{\zbl}/2$),
then $Y_{N_1}$ decreases. The decrease in the decay width
results in a corresponding decrease of $Y_{N_1}$ as expected.
However we find no visible change in $Y_{N_1}$ when $\mdm$
goes from 1 MeV to 100 MeV. This is because in both of these cases
$M_{\zbl} \gg 2 \mdm$, therefore the decay
width $\Gamma_{Z_{\rm BL} \rightarrow N_1 N_1}$ and hence $Y_{N_1}$
is practically insensitive to $\mdm$ ($\mdm=$ 1 MeV to 100 MeV).\\
Finally, the effect of the variation of gauge boson mass $M_{\zbl}$
on $Y_{\zbl}$ and $Y_{N_1}$ is shown in Fig. \ref{plot3}(d).
The increase in $M_{Z_{\rm BL}}$ results in the decrease
of $\Gamma_{H \rightarrow Z_{\rm BL}Z_{\rm BL}}$ and an
increase of $\Gamma_{Z_{\rm BL} \rightarrow all}$, 
which is manifested through a smaller rise and a
faster fall of $Y_{Z_{\rm BL}}$. This nature of $Y_{\zbl}$ is
corroborated in the plot as well, where $M_{\zbl}$ varies
from 10 GeV (red line) to 100 GeV (green line).
On the other hand the quantity $Y_{N_1}$ follows the
evolution of $Y_{Z_{\rm BL}}$ in the usual way as discussed earlier.
Since all these cases were shown to demonstrate the validity
of the coupled Boltzmann equations, for simplicity the active
sterile mixing angle ($\alpha$) is set to zero.
\subsection{Solution of the complete Boltzmann equation(s)
with all production and decay channels}
\label{sec_BE_full}
\begin{figure}[h!]
\centering
\subfigure[Variation with different $\gbl$ values]
{\includegraphics[scale=0.31,angle=-90]{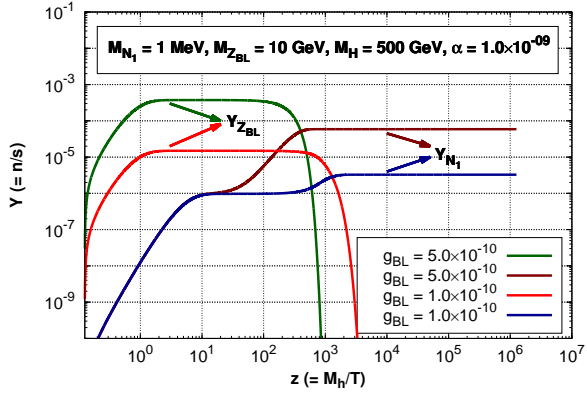}}
\subfigure[Variation with different $M_H$ values]
{\includegraphics[scale=0.31,angle=-90]{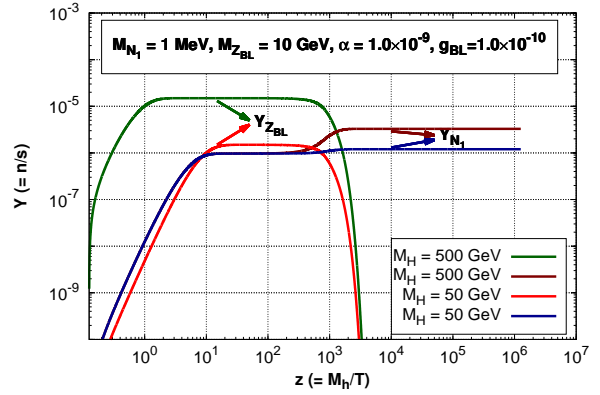}}\\
\subfigure[Variation with different $\alpha$ values]
{\includegraphics[scale=0.31,angle=-90]{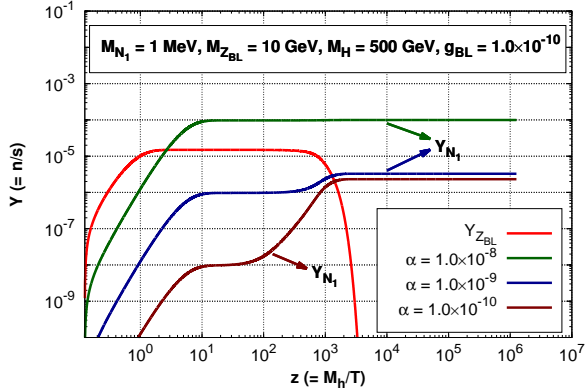}}
\subfigure[Variation with different $\mzbl$ values]
{\includegraphics[scale=0.31,angle=-90]{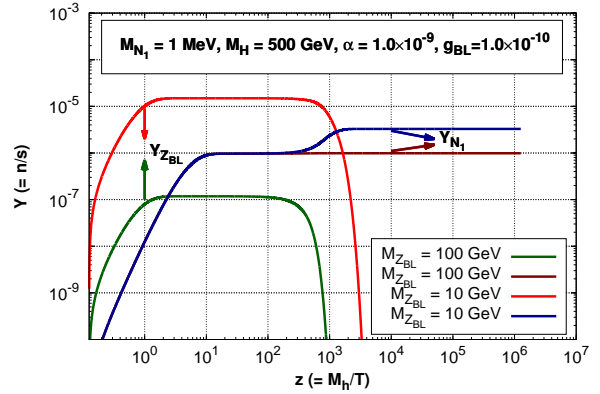}}
\caption{Comparison of $Z_{\rm BL}$ and $N_1$ comoving number density
with respect to different sets of chosen parameters.}
\label{plot4}
\end{figure}

In the previous section, we demonstrated the validity of the coupled
set of Boltzmann equations needed to solve for the relic abundance
of the sterile neutrino dark matter $N_1$. For simplicity, we assumed
that the only production channel of $N_1$ is the decay from $\zbl$.
However in general, all the possible production modes of $N_1$ including decays
as well as annihilations of SM and BSM particles, as given in Eq. (\ref{bltz2}),
have to be taken into account. Therefore, the active-sterile mixing angle
is now nonzero. The noticeable feature when the active sterile mixing
is nonzero is the production of $N_1$ from the decay of
$W^{\pm}$ bosons ($W^\pm \rightarrow e^\pm N_1$).
It may a priori seem that due to small value of active-sterile
mixing angle the contribution from the decay of $W^\pm$ will
be negligible, but we have to remember that in this
non-thermal scenario, the extra gauge coupling ($\gbl$)
is also required to be very small ($\sim 10^{-10}$),
and hence the production of $N_1$ from the decay of $\zbl$ may
also compete with the former. We will show this quantitatively
later. Also note the decay of $W^{\pm}$ is solely governed by
the active-sterile mixing angle $\alpha$ and does not depend on
U(1)$_{\rm B-L}$ gauge coupling $\gbl$, hence if $\gbl$ is
made very low, the {\it only} dominant production channel of
$N_1$ will be from $W^\pm$ decay. In Fig. \ref{plot4}((a)-(d)) we
show the variation $Y_{N_1}$ and $Y_{\zbl}$ with respect to different
sets of independent parameters as before.
\\
Interesting feature of Fig. \ref{plot4}((a)-(d)) when contrasted
with Fig. \ref{plot3}((a)-(d)) is the existence of a ``double plateau''.
The reason behind this is the presence of another the production mode of $N_1$
from $W^\pm$ decay which was neglected in previous section for simplicity.
The onset of $N_1$ production as seen in these plots is a little early
than those seen in Fig.\ref{plot3}. The initial onset here is due to
the presence of $W^\pm$ decay and is independent of $\gbl$, $M_{\zbl}$ and $M_H$
(Eq. (\ref{wn1e})), and only depends on $\alpha$ which is the sole parameter
that controls the $W^\pm \rightarrow e^\pm N_1$ decay. The first plateau occurs
when the number density of $W$ boson begins to fall and hence there is a
decreased rate of production of $N_1$. However it again begins to rise
sharply when the production from $\zbl$ starts to dominate. Then as before,
we can see by comparing with the accompanying $Y_{\zbl}$ lines that
the second plateau results when the $\zbl$ number density starts
to deplete. It is to be noted that this ``two plateau'' feature will
be visible only when the production from $W$ and $\zbl$ are comparable
to each other at some point of $z(=\frac{M_h}{T})$. If either one of
them remains dominant for all $z$, then it will result in single plateau
like feature. For example, the green solid line in plot (c) of Fig. \ref{plot4}
has only a single plateau. This is because, due to a very high value of $\alpha$
the decay channels of $W^{\pm}$ remain the most dominant production
mode of $N_1$ for all $z$. The corresponding variation of $\zbl$ number
density is also shown by red solid line and since $\alpha$ has no effect
on the production and decay channels of $\zbl$ we find no variation
of $Y_{\zbl}$ with $\alpha$. However the variation of $Y_{N_1}$
is different for different values of $\alpha$, as the active-sterile
mixing angle $\alpha$ controls the production mode of $N_1$ from $W^{\pm}$ decay.  
\section{Relic Density of Sterile Neutrino Dark Matter ($N_1$)}
\label{sec_RD}
\begin{figure}[h!]
\centering
\subfigure[$W$ contribution to $\Omega_{N_1}h^2$ $\sim$
70\% while $\zbl$ contribution $\sim$ 30\%]
{\includegraphics[scale=0.3,angle=-90]{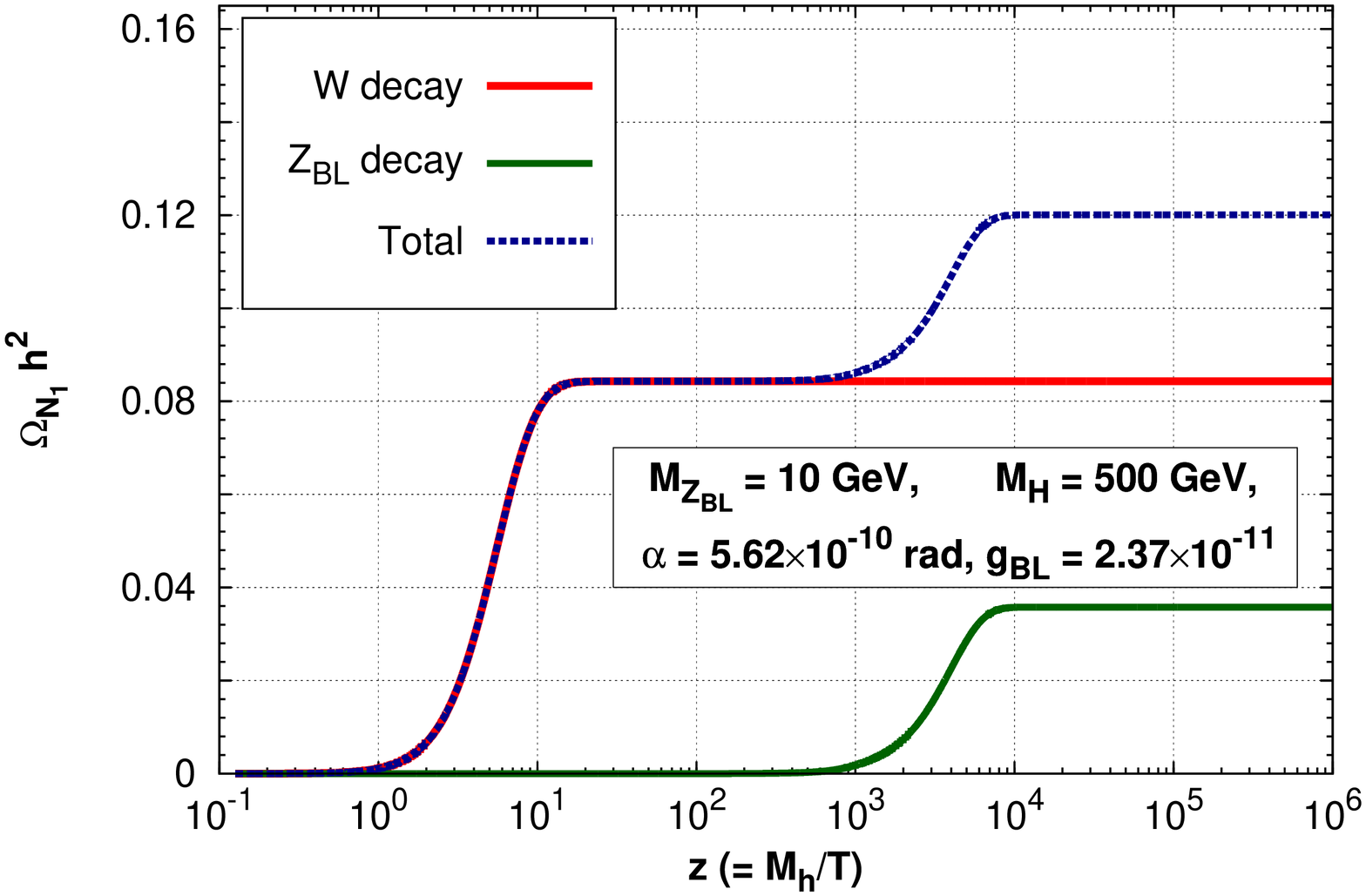}}~~
\subfigure[$W$ contribution to $\Omega_{N_1}h^2$ $\sim$  $\zbl$ contribution]
{\includegraphics[scale=0.3,angle=-90]{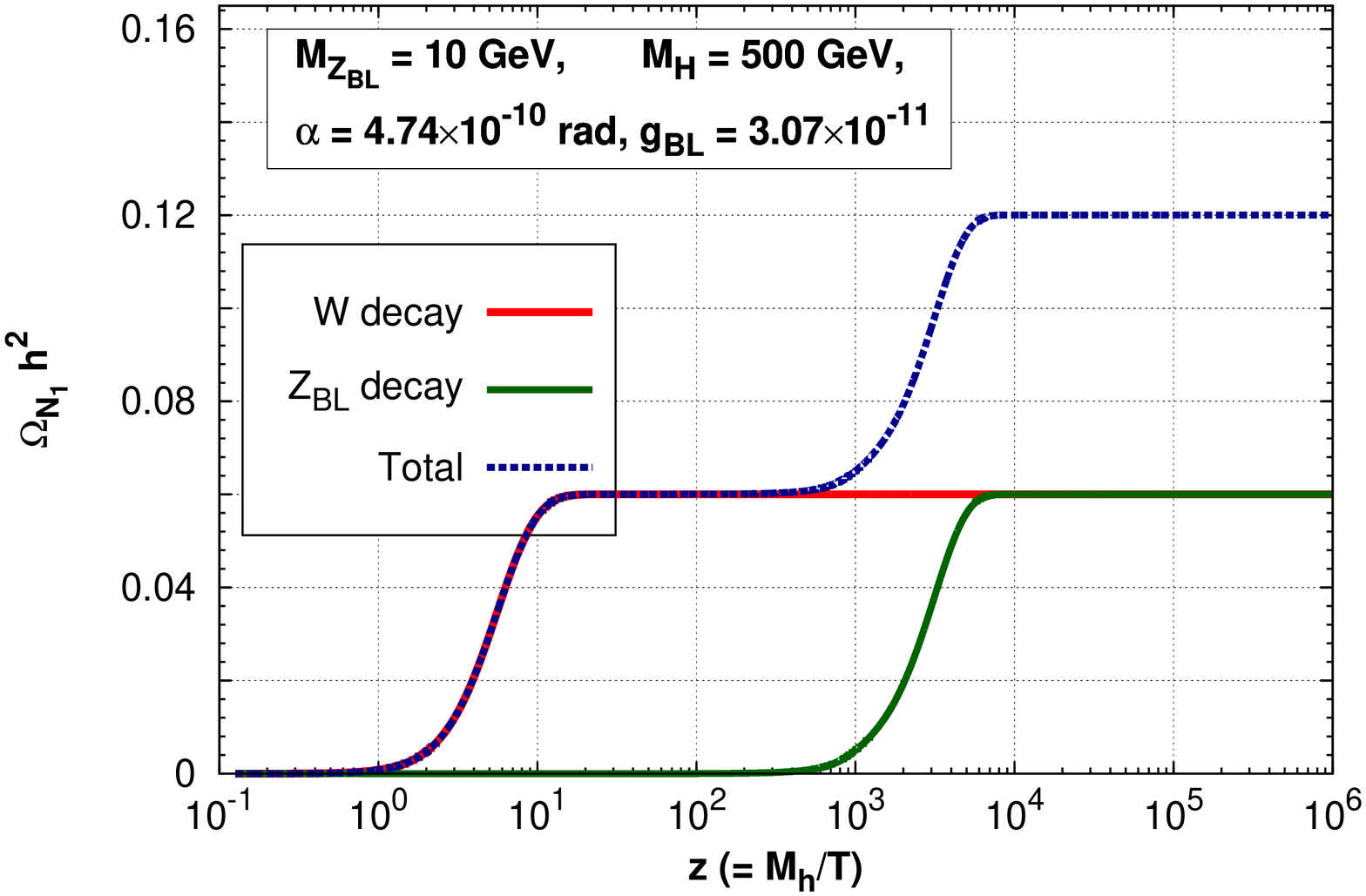}}\\
\subfigure[$W$ contribution to $\Omega_{N_1}h^2$ $\sim$ 30\%
while $\zbl$ contribution $\sim$ 70\%]
{\includegraphics[scale=0.3,angle=-90]{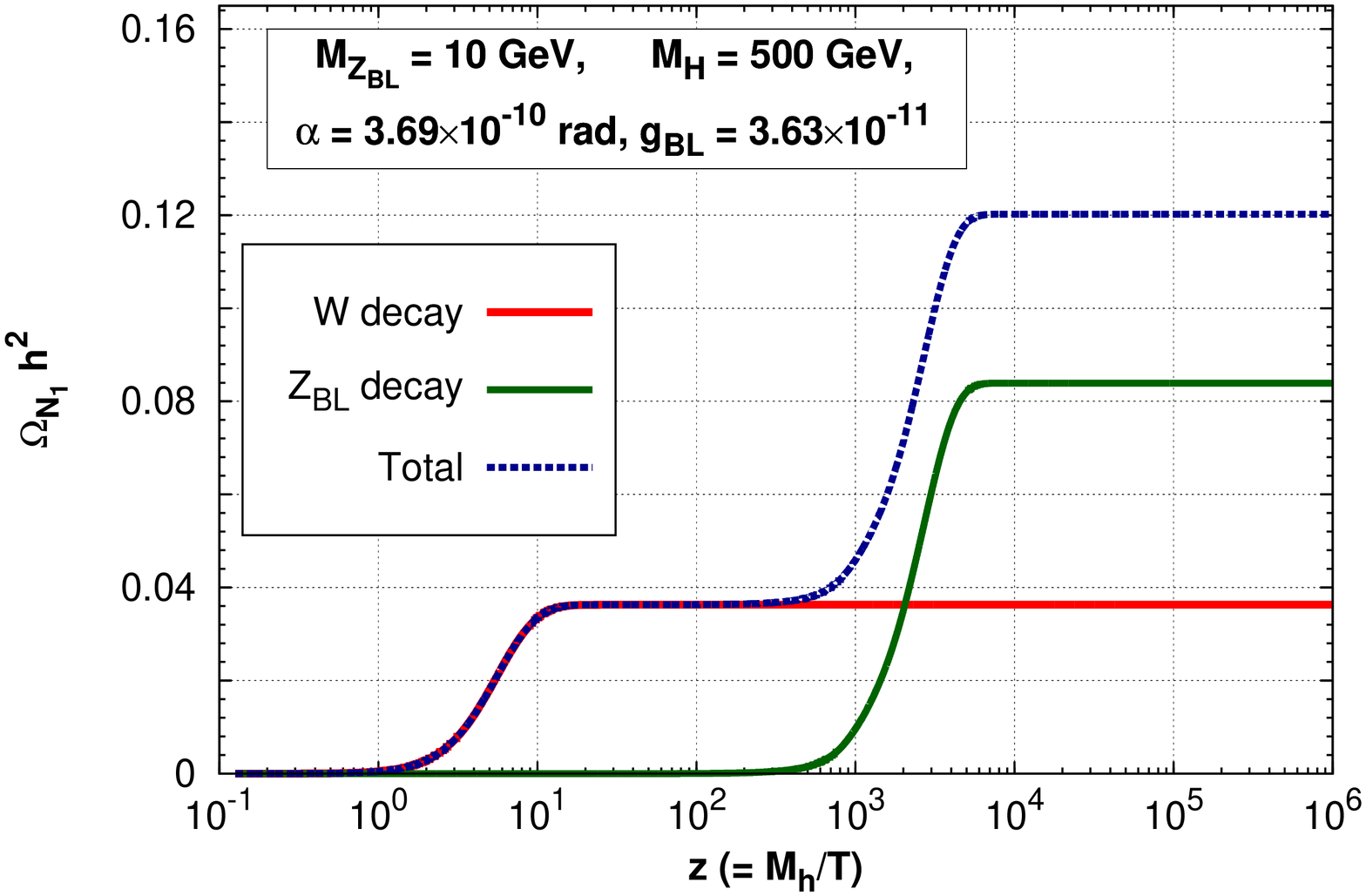}}
\caption{Relic abundance of $N_1$ as function of $z$ along with the relative
contributions of $W^{\pm}$ and $\zbl$ decay channels. All the plots are drawn for
$\mdm=1$ MeV.}
\label{plot5}
\end{figure}

In order to compute the relic abundance ($\Omega_{N_1} h^2$)
of the lightest sterile neutrino ($N_1$)
we need to find the value of its comoving number density ($Y_{N_1}$) at the
present epoch ($T=T_0$, $T_0\sim2.73$ K). The value of $Y_{N_1}(T_0)$ can be
obtained by solving the two coupled Boltzmann equations
(Eqs. (\ref{bltz1}, \ref{bltz2}), which we have
discussed elaborately in Section \ref{sec_BE}. The expression of
$\Omega_{N_1} h^2$ in terms of $Y_{N_1}(T_0)$
is given by \cite{hep-ph/9704361},
\begin{eqnarray}
\Omega_{N_1} h^2 = 2.755\times 10^8
\left(\frac{\mdm}{\rm GeV}\right) Y_{N_1}(T_0)\,\,.
\end{eqnarray}
In this work, we take all decay and annihilation channels of both SM as well as
BSM particles for the production of $N_1$. In Fig. \ref{plot5} we show the
relative contributions to $\Omega_{N_1} h^2$ from $W^{\pm}$ (red solid line)
and $\zbl$ decay (green solid line) for some chosen sets of model parameters.
The total relic abundance of $N_1$ is also shown by the blue solid line. For some
combinations of model parameters we find $W^{\pm}$
decay can be the leading production channel of $N_1$
(plot (a)) while for some others it can be the subleading one (plot (c)).
However in all three plots (a-c) of Fig. \ref{plot5}, the total relic density of
$N_1$ has the saturation value $\sim 0.12$ which is in conformity with
the value of dark matter relic density measured by the satellite borne
experiment Planck. 
In plot (b) of Fig. \ref{plot5} we show a situation when
the relative contributions to $\Omega_{N_1} h^2$ from both
$W^{\pm}$ and $\zbl$ decays are equal. In this case
we have adopted the following values of relevant model parameters:
$\gbl = 3.07\times 10^{-11}$, $\alpha = 4.74\times 10^{-10}$ rad,
$\mzbl=10$ GeV and $M_H=500$ GeV. For this set of model
parameters we have also listed the fractional
contributions to $\Omega_{N_1} h^2$ arising
from all the possible decay and annihilation channels
in Table \ref{tab1}, \ref{tab2}.

\begin{table}[h!]
\begin{center}
\begin{tabular} {||c||c||}
\hline
\hline
Decay Channel & Fractional contribution to $\Omega_{\rm DM}h^2$\\
\hline
\hline
$W^{\pm}$ & 0.5000\\
\hline
$Z_{\rm BL}$ & 0.4999\\
\hline
$H$ & $0.2590\times 10^{-10}$\\
\hline
$h$ & $0.1177\times 10^{-11}$\\
\hline
$Z$ & $0.6276\times 10^{-19}$\\
\hline
\hline
\end{tabular}
\end{center}
\caption{Fractional contributions of different production
processes of $N_1$ through decay for $\gbl = 3.07\times 10^{-11}$,
$\alpha = 4.74\times 10^{-10}$ rad, $\mzbl=10$ GeV, 
$M_H=500$ GeV and $\mdm=1$ MeV.}
\label{tab1}
\end{table}

\begin{table}[h!]
\begin{center}
\begin{tabular} {||c||c||}
\hline
\hline
Annihilation Channel & Fractional contribution to $\Omega_{\rm DM}h^2$\\
\hline
\hline
$t\bar{t}$ & $0.3745\times 10^{-12}$\\
\hline
$hh$ & $0.1650\times 10^{-13}$\\
\hline
$W^+ W^-$ & $0.3606\times 10^{-14}$\\
\hline
$ZZ$ & $0.3562\times 10^{-14}$\\
\hline
$HH$ & $0.4403\times 10^{-19}$\\
\hline
$Z_{\rm BL} Z_{\rm BL}$ & $0.4515\times 10^{-30}$ \\
\hline
$N_2 \bar{N}_2$ & $0.1987\times 10^{-35}$ \\
\hline
$N_3 \bar{N}_3$ & $0.1987\times 10^{-35}$ \\
\hline
\hline
\end{tabular}
\end{center}
\caption{Fractional contributions of different production
processes of $N_1$ through annihilation for $\gbl = 3.07\times 10^{-11}$,
$\alpha = 4.74\times 10^{-10}$ rad, $\mzbl=10$ GeV, 
$M_H=500$ GeV and $\mdm=1$ MeV.}
\label{tab2}
\end{table}
From Table \ref{tab1} it is seen that for the small values
of $\gbl \sim 10^{-11}$ and $\alpha \sim 10^{-10}$ (which are
required for the non-thermality of $N_1$) the
contributions of other production channels of $N_1$
through the decays of $Z$, $H$ and $h$ are negligible. 
Similarly, Table \ref{tab2} shows that within this adopted
ranges of model parameters the annihilation processes of SM as well as
BSM particles do not contribute significantly to the production
of sterile neutrino dark matter $N_1$ and hence we can safely consider
the decays of $W^{\pm}$ and $\zbl$ as the two most efficient production
mechanisms of $N_1$.

\begin{figure}[h!]
\centering
\includegraphics[height=6cm,width=9cm]{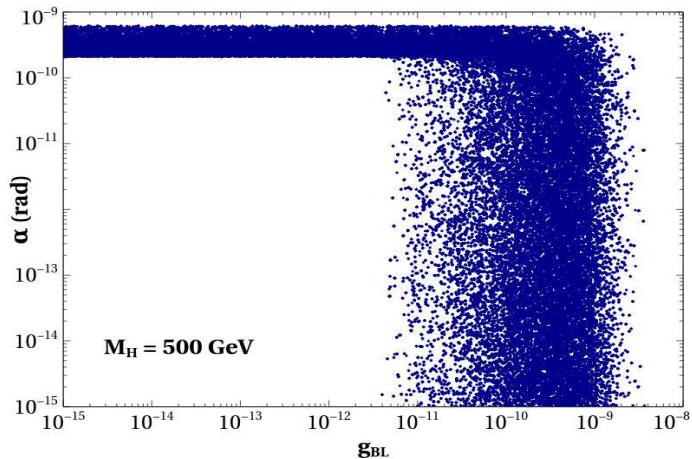}
\caption{Allowed region in $\gbl$ Vs $\alpha$ plane satisfying the relic density
criteria.}
\label{plot6}
\end{figure}

In Fig. \ref{plot6}, we plot the allowed values of $\rm B-L$ gauge coupling
$\gbl$ and active sterile mixing angle $\alpha$ which satisfy the relic density
criteria ($0.1172 \leq \Omega_{N_1} h^2 \leq 0.1226$) \cite{1502.01589}.
During the computation of Fig. \ref{plot6} we have varied the relevant
parameters in the following range.
\begin{eqnarray}
\begin{array}{cccccc}
0.1 \text{ GeV}& \le & \mzbl & \le & 250\text{ GeV}\,\, ,\\
1.2\text{ MeV}&\le & \mdm & \le& 10\text{ MeV}\,\, ,\\
10^{-8}&\le & \gbl & \le& 10^{-15}\,\, ,\\
10^{-7}\text{ rad}&\le & \alpha & \le& 10^{-17}\text{ rad}\,\, ,
\label{para-ranges}
\end{array}
\end{eqnarray}
and we have kept the mass of the extra Higgs boson $H$ fixed at 500 GeV.
From Fig. \ref{plot6}, we see that for very small values of the extra
gauge coupling $\gbl$ ($\sim 10^{-12}\,\,{\rm to}\,\,10^{-15}$)
the relic density condition of $N_1$ is always
satisfied for a active-sterile mixing angle $\alpha \sim 10^{-10}$ rad.
This is expected since for very small values of $\gbl$ the production
of $N_1$ from the decay of $\zbl$ is highly suppressed and in this situation
decay of $W^\pm$ becomes the principle production channel,
since the latter is {\it not} suppressed by the extra gauge coupling
(see Eqs. ({\ref{wn1e}}), (\ref{zbln1n1})). Earlier works about the non-thermal
production of sterile neutrino have not touched upon this point in detail
(production of $N_1$ from $W^{\pm}$ decay), since most of the previous authors
have ignored the $N_1$ production mode from $W^{\pm}$ decay in their
works by assuming extremely small values of active-sterile mixing angle $\alpha$.
However, such an assumption needs careful attention when other
couplings in the theory can also be very small.
On the other hand for the higher values of
$\gbl$ ($\sim 10^{-9}\,\,{\rm to}\,\,10^{-11}$)
the decay of $\zbl$ becomes the dominant
contributor and hence in this case small values of $\alpha$ are required
to suppress the contribution of $W^{\pm}$ decay to $\Omega_{N_1} h^2$
such that the total relic density of $N_1$ lies within
the range prescribed by the Planck experiment. Since the production
of sterile neutrino from $W^\pm$ decay depends only on the mixing angle
$\alpha$, in Fig. \ref{plot6} we get only a narrow band of $\alpha$ which
satisfies the relic density of $N_1$ (for small values of $\gbl$).
But when $\zbl$ is the main production
channel of $N_1$ then for a given $\gbl$ and $\alpha$ we can make $N_1$
to satisfy the relic density by adjusting the $\zbl$ mass. Hence we get a
relatively wider band of allowed values of $\gbl$ for a fixed $\alpha$. 
For the chosen mass range of sterile neutrino (i.e. $\mathcal{O}$(MeV)),
from Fig. \ref{plot6} we find that the maximum allowed value of $\alpha$
is $\sim 10^{-10}$ rad. Such a sterile neutrino is free from all the
constraints arising from X-ray and BBN as seen from
Fig. 1 of Ref. \cite{1512.02751}.


\begin{figure}[h!]
\centering
\includegraphics[height=6cm,width=9cm]{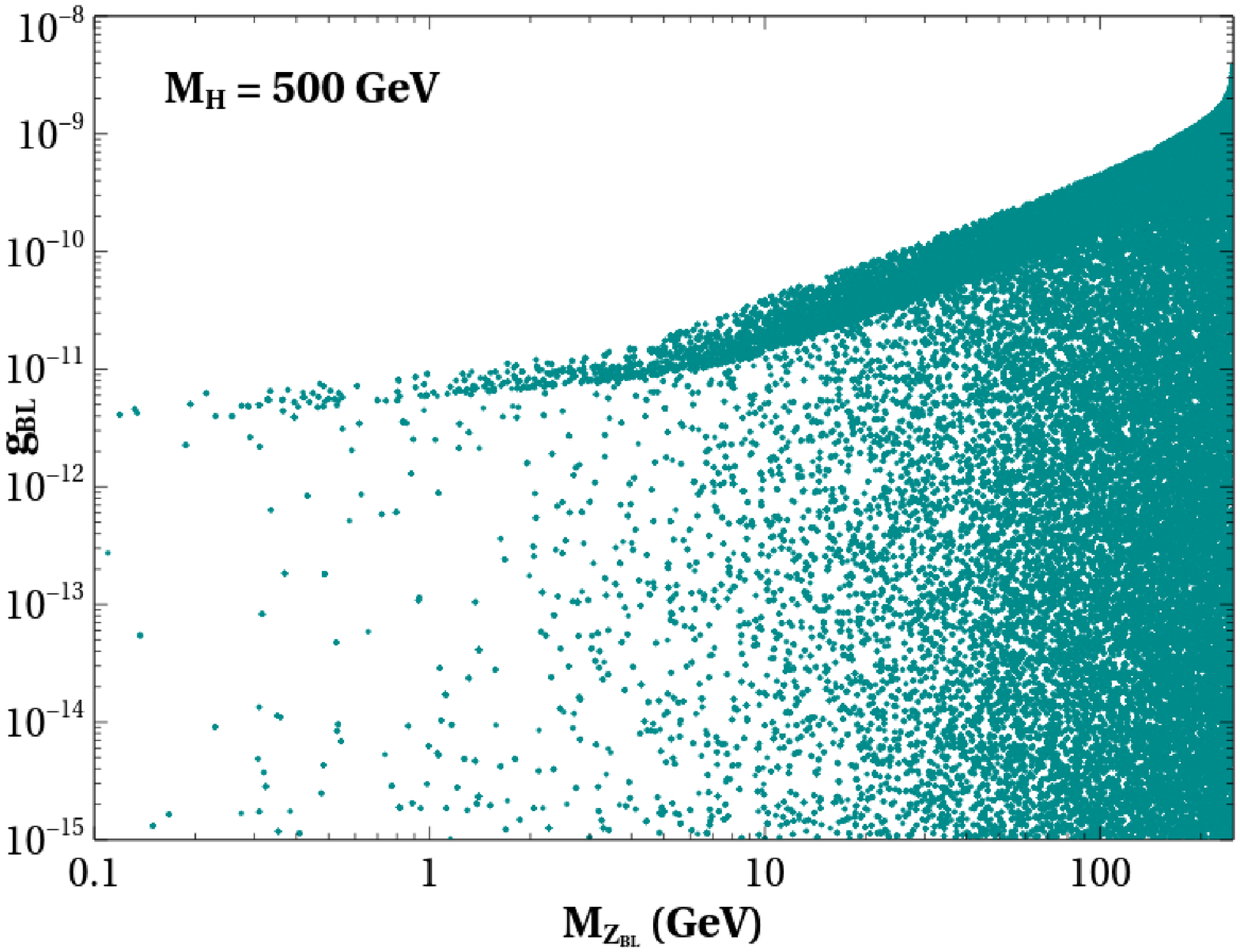}
\caption{Allowed region in $\mzbl$ Vs $\gbl$ plane satisfying the relic density
criteria.}
\label{plot7}
\end{figure}

The allowed region in $\mzbl-\gbl$ plane is shown in Fig.\,\,\ref{plot7}.
Like the previous plot in Fig. \ref{plot6}, here also all the points in
$\mzbl-\gbl$ plane produce the correct relic density of $N_1$
($0.1172 \leq \Omega_{N_1} h^2 \leq 0.1226$). In this case we
have also varied the other relevant parameters ($\mdm$, $\alpha$)
in the range given in Eq. (\ref{para-ranges}). From this figure it
is seen that the allowed values of $\gbl$ increases with $\mzbl$.
This nature of $\mzbl-\gbl$ plane can be explained in the following way.
We know that the contribution of $\zbl$ to $Y_{N_1}$ depends on both
$\Gamma_{\zbl\rightarrow N_1 N_1}$ and $Y_{\zbl}$ (see Eq. (\ref{bltz2}))
where the latter quantity increases with
$\Gamma_{H\rightarrow \zbl \zbl}$ (Eq. (\ref{bltz1}))
as the extra gauge bosons $\zbl$ are produced mainly from the decay of $H$.
However the decay width $\Gamma_{H\rightarrow \zbl \zbl}$ is suppressed by
$\mzbl^{-2}$ and thereby reducing the comoving number density of $\zbl$
with its mass (see Eq. (\ref{Hzblzbl}) and Fig. \ref{plot4}(d)).
In order to keep the contribution to $\Omega_{N_1} h^2$
arising from $\zbl$ decay unaltered, this decrement in
$Y_{\zbl}$ must be compensated by a corresponding increment in
$\Gamma_{\zbl\rightarrow N_1 N_1}$ which is
proportional to $\gbl^2$ (Eq. (\ref{zbln1n1})).
Hence with an increase in $M_{\zbl}$, $\gbl$
should also increase to satisfy the relic density constraint.

Simulations using the standard $\Lambda \rm CDM$ cosmology requires
that most of the dark matter candidates should be {\it cold}
to satisfy constraints from the structure formation
\cite{astro-ph/9707285, 0906.4340}.
In our case, to get an idea about the {\it coldness} of the
sterile neutrino dark matter we try to compute its free-steaming
length defined by \cite{0812.0010}:
\begin{eqnarray*}
\lambda_{\rm fs} =\int_{t_{in}}^{t_0}\dfrac{\langle v(t) \rangle}{a(t)}\,dt\,,
\end{eqnarray*}
where $t_{in}$ is the initial time, $t_0$ is the present time,
$v(t)$ is the mean velocity of the dark matter, and $a(t)$ is
the scale factor of the Universe. Following Ref. \cite{1306.3996},
the hot, cold and warm dark matters are classified as:
\begin{eqnarray*}
\rm Cold\,\, Dark \, \,Matter \, \, (CDM) &:& \lambda_{\rm fs} < 0.01 \, \rm Mpc \\
\rm Warm\,\, Dark \, \,Matter \, \, (WDM) &:& 0.01 \, \rm Mpc 
< \lambda_{\rm fs} < 0.1 \, \rm Mpc \\
\rm Hot\,\, Dark \, \,Matter \, \, (HDM) &:& \lambda_{\rm fs} > 0.1 \rm \, Mpc
\end{eqnarray*}
For the case where both $W^\pm$ and $\zbl$ contribute equally
to the final dark matter relic density (Fig. \ref{plot5}b), we
have calculated the free-streaming length for dark matter
produced from $W^\pm$ as well as $\zbl$ decay separately. In both
the cases we find that $\lambda_{\rm fs}^{W^\pm}\sim 0.003$ Mpc and
$\lambda_{\rm fs}^{\zbl}\sim 0.007\,\,{\rm Mpc}$. Hence following the above
classification of hot, cold and warm dark matter, we find
that all of our sterile neutrino is {\it cold} and thus
satisfying the structure formation constraints.

\section{A possible way of detecting the sterile neutrino Dark Matter}
\label{integral}
In 2003 INTEGRAL/SPI \cite{astro-ph/0309442} of ESA  observed
an emission line at an energy of 511 keV mostly
from the galactic bulge. Recently, it has been
reported that the measured flux from the galactic bulge
by INTEGRAL/SPI is $\Phi_{511}^{\rm exp} =(0.96 \pm 0.07)
\times 10^{-3} {\rm ph}\,{\rm cm}^{-2}\,{\rm s}^{-1}$
at 56$\sigma$ significance \cite{1512.00325}. A possible source of this
line is assumed to be the annihilation of electron and positron
in the galactic core. Inspite of some astrophysical processes
explaining the origin of the line \cite{astro-ph/0506026},
the sources of the galactic positrons are not
clear yet. Hence a series of possible explanations
have been reported in last ten years involving positrons originating
from a decaying \cite{1206.3076, hep-ph/0607076} or
annihilating \cite{astro-ph/0702587, astro-ph/0311150} dark matter.
For a brief review of earlier works trying to explain 511 keV
line see \cite{1009.4620}.
Recently the authors of Ref.\,\,\cite{1602.01114} have shown that the
explanation of this anomalous emission line is not possible
from the annihilation of thermal dark matter (WIMP) due to conflict with the
latest cosmological data and they have preferred a
non-thermal origin of dark matter for explaining this
long standing puzzle. Earlier people have tried
to explain this INTEGRAL anomaly from the decay
of light sterile neutrino dark matter \cite{hep-ph/0402178, 0804.0336}.
Here, in the case of sterile neutrino dark matter
in a non-thermal setting, we have also found that such an
explanation is indeed possible. The decaying dark matter
scenarios however require a more cuspy density profile
than the annihilation models \cite{astro-ph/0507142}. The seed mechanism behind
this 511 keV emission line is the decay of sterile neutrino
($N_{1}$) into a e$^\pm$ pair and an active neutrino.
The $e^{\pm}$ pair thus produced, get slowed down
to non relativistic velocities due to several energy
loss mechanisms within the galactic bulge \cite{astro-ph/0309686} and
thereby producing 511 keV {\it{gamma-line}} from their pair
annihilation. The mass of the sterile neutrino
favourable to explain this signal is $\sim$ 1-10 MeV \cite{astro-ph/0309686}.
In the present U(1)$_{\rm B-L}$ model, there are six possible
Feynman diagrams contributing to this three body decay of
which those which are mediated by $\zbl$, $h$ and $H$ are
sub dominant due to the suppression by the very low value of
U(1)$_{\rm B-L}$ gauge coupling $\gbl$.
Therefore, we have used the remaining three diagrams
i.e. those mediated by $Z$ and $W^\pm$ bosons to calculate the
three body decay width. The expression of
matrix amplitude squared and corresponding decay width
$\Gamma_{N_1 \rightarrow e^{\pm} \nu}$ is given in Appendix \ref{gamman1enu}.
A more simpler analytical expression (using some approximation)
for this three body decay width can be found in Ref. \cite{hep-ph/0402178}.
The expression for the gamma ray flux obtained from the
galactic bulge due the decay $N_1 \rightarrow e^\pm
\, \nu$ is given by \cite{hep-ph/0402178}:
\begin{eqnarray}
\Phi_{511}^{theory} &=& 2 \, \frac{1}{4\pi}
\frac{\Gamma_{N_1 \rightarrow e^\pm \, \nu}}{\mdm}
\dfrac{\int_{\Delta \Omega}\,\int_{l.o.s}
\,\rho_{\rm DM}(r(s,\,\Omega))\,ds\,d\Omega}{\int_{\Delta \Omega} d\Omega}\,.
\label{511}
\end{eqnarray}
Here, $\Gamma_{N_1 \rightarrow e^\pm \, \nu}$ is the
decay width of $N_1 \rightarrow e^\pm \, \nu$ and
$\rho_{\rm DM}(s,\,\Delta \Omega)$ is the dark matter density profile
in the galaxy. During our analysis, we have taken Einasto profile \cite{1012.4515}
with $\alpha_{einasto} = 0.17$ for the computation of gamma-ray flux.
The angular integration over the solid angle $\Delta \Omega$ is performed
within the 2$^\circ$ angular resolution of the spectrometer while
the spacial integration is over the line of sight (l.o.s) distance
of galactic bulge from the position of solar system. The extra factor of 2 appearing in Eq. (\ref{511}) is due to the production of two photons
per decay of $N_1$.

\begin{figure}[h!]
\centering
\includegraphics[height=7cm,width=9cm]{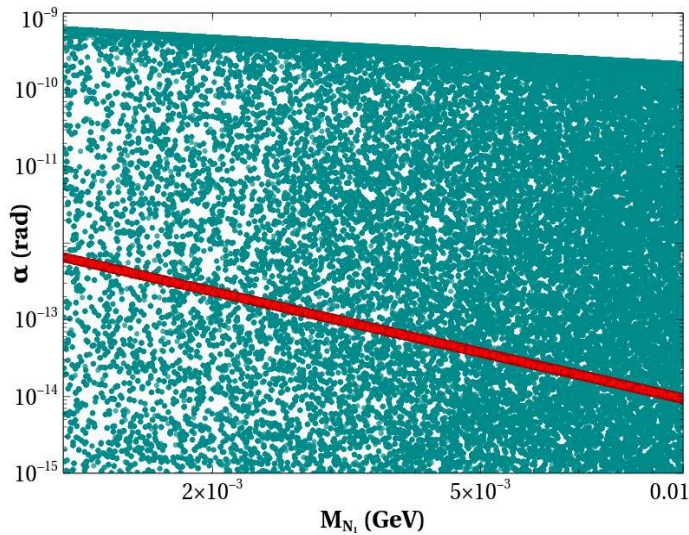}
\caption{Values of $\mdm$ and active-sterile mixing angle $\alpha$ allowed
by the relic density of $N_1$ ($0.1172 \leq \Omega_{N_1}h^2 \leq 0.1226$),
are shown by the dark cyan points. The points lying within the red coloured
band reproduced the flux observed by INTEGRAL/SPI.}
\label{intg}
\end{figure}

Using Eq. (\ref{511}) we have computed the photon flux for
different values of $\mdm$ and $\alpha$. In Fig. (\ref{intg})
the red band shows the correct combination of $\mdm$ and $\alpha$
which is needed to explain the INTEGRAL observed flux. The dark cyan
region is for those values of $\mdm$ and $\alpha$ which satisfies
only the relic density constraint of $N_1$. From Fig. \ref{intg},
we can see that in the chosen range of $\mdm$ ($\sim$ 1 - 10 MeV)
the active-sterile mixing angle $\alpha$ required to explain
{\it both} the relic density as well as the INTEGRAL anomaly
is $\sim 10^{-12}-10^{-14}$ rad.
\section{Conclusion}
\label{conclu}
In this work we have shown that non-thermal sterile neutrino in
$\ubl$ model can be a viable dark matter candidate. But the formalism
developed here is in general applicable to any U(1)$_{X}$ extension of
the Standard Model. Any such model trying to describe a non-thermal
dark matter scenario (through IR Freeze-in) will have in general a
very weakly coupled $Z^\prime$ as well as a feebly coupled dark matter
candidate. Under such circumstances (i.e. when the mother particle
responsible for most of the production of the dark matter has itself
gone out of thermal equilibrium), we have shown how to solve a set
of coupled Boltzmann equations to calculate the final relic density.
We have seen that the sterile neutrinos are mostly produced from the
decay of $\zbl$ and $W^\pm$. We have also shown that though the contribution
from $W^\pm$ was neglected in the previous works (under the assumption
of smallness of the active-sterile mixing angle), it can actually be
sizeable  (even dominating over the production from $\zbl$ decay for
some values of $\alpha$ and $\gbl$) depending on the parameter space
we are focussing on. Note that for generic values of $\alpha$ we use
in this work, one of the active neutrinos has to be very light. However
this is allowed by the data available from the various present day
neutrino experiments. Finally for completeness, we have also checked
that such an $\mathcal{O}$(MeV) mass non-thermal sterile neutrino can
explain the 511 keV line observed by INTEGRAL/SPI. The $\alpha$ required
to explain this signal falls in the region where the dark matter production
is mostly dominated by $\zbl$ decay. The decay width required has a
corresponding life time $\sim 10^{25}$ s, which much larger than the
present age of the Universe ($\sim 10^{17}$ s).
\section{Acknowledgement}
Authors would like to thank Raj Gandhi, Alexander Merle,
Sourav Mitra, Tirthankar Roy Choudhury, Osamu Seto and Bibhushan Shakya
for many useful suggestions and discussions.
Authors would also like to thank Mehedi Masud, Avirup Shaw and
Taushif Ahmed for helping out with various aspects of numerical calculations.
Authors also acknowledge Department of Atomic Energy (DAE), Govt. of
India for their financial assistance and the cluster computing facility at HRI 
(http://cluster.hri.res.in).
\appendix
\section{Analytical expressions for cross sections and decay widths}
\subsection{Production processes of $N_1$ from the decays of
SM and BSM particles}
\label{decay_channel}
In this section, we give the expressions of all the relevant decay widths
which are needed to solve the coupled Boltzmann equation for $N_1$
(Eqs. (\ref{bltz2})).
\begin{eqnarray}
\Gamma(W^{+} \rightarrow N_{1}\,e^+) &=& \frac{2\,(2M_{W}^{4}-(M_{e}^{2}-
M_{N_{1}}^{2})^{2}-(M_e^{2}+M_{N_{1}}^{2})M_{W}^{2})
\sin^{2}\alpha}{3v^2}\times \nonumber \\
&& \frac{\sqrt{1-\left( \frac{M_e+M_{N_{1}}}{M_{W}}\right)^{2}}
\sqrt{1-\left( \frac{M_e-M_{N_{1}}}{M_{W}}\right)^{2}}}{16\pi M_{W}} \,,
\label{wn1e}\\
\Gamma(Z\rightarrow N_{1} \bar{N_1})&=& \frac{M_{Z}^3\,
\sin^4\alpha}{24\pi\,v^2}\left(1-\frac{4M_{N_{1}}^2}{M_{Z}^2}\right)^{3/2}\,,
\label{zn1n1}\\
\Gamma(Z_{\rm BL}\rightarrow N_1 \, \bar{N_1})&=&
\frac{M_{Z_{\rm BL}}}{24\pi}\,((\sin^2\alpha-\cos^2\alpha)^2
\,g_{\rm BL}^2)\left(1-\frac{4M_{N_{1}}^2}{M_{Z_{\rm BL}}^2}\right)^{3/2}\,,
\label{zbln1n1}\\
\Gamma(H \rightarrow N_1 \bar{N_1})&=& \frac{g_{HN_1N_1}^2\,M_H}
{16\pi}\left(1-\frac{4M_{N_{1}}^2}{M_{H}^2}\right)^{3/2}\,,
\label{Hn1n1}\\
\Gamma(h \rightarrow N_1 \bar{N_1})&=& \frac{g_{hN_1N_1}^2\,M_h}
{16\pi}\left(1-\frac{4M_{N_{1}}^2}{M_{h}^2}\right)^{3/2}\,,
\label{hn1n1}
\end{eqnarray}

where $M_x$ denotes the mass of particle $x$, $\alpha$ is the active-sterile
mixing angle of first generation (i.e. mixing of $\nu_1$ with $N_1$) while
$g_{H N_1 N_1}$ and $g_{h N_1 N_1}$ are vertex factors
corresponding to the vertices $H N_1 N_1$ and $h N_1 N_1$ respectively.
With respect to our chosen set of independent parameters,
these vertex factors are given as:
\begin{eqnarray}
g_{H N_1 N_1} &=&
2\,\cos\alpha \left(\sin\theta\sin\alpha 
\frac{\sqrt{M_{N_1}m_{\nu_1}}}{v}-\cos\theta\cos\alpha\,
\frac{g_{\rm BL}\,M_{N_1}}{M_{Z_{\rm BL}}}\right)\,,\nonumber \\
g_{h N_1 N_1}&=&2\,\cos\alpha\left(\cos\theta\sin\alpha
\frac{\sqrt{M_{N_1}m_{\nu_1}}}{v}+\sin\theta\cos\alpha\,
\frac{g_{\rm BL}\,M_{N_1}}{M_{Z_{\rm BL}}}\right)\,.
\end{eqnarray}

\subsection{Production processes of $N_1$ from annihilation}
\label{annihilation_channel}
In this section, we present the expressions of all the relevant
annihilation cross sections i.e. the production processes of
$N_1$ through the annihilations of SM as well as BSM particles.
In all the expressions given below, $M_X$ and $\Gamma_X$ denote
the mass and total decay width of the particle X while $g_{ijk}$ denotes
the coupling of the vertex involving fields $i,j,k$. Further, $\sqrt{s}$ is
the center of mass energy of a particular annihilation process.
All the annihilation cross sections given below are written in
terms of our chosen set of independent parameters.\\

$\underline{\mathbf{W^+ \, W^- \rightarrow N_1 \bar{N_1}}}$\\

In this annihilation process, three $s$-channel diagrams mediated by $h$, $H$
and $Z$ and one electron mediated $t$-channel diagram are possible. However,
$Z$ boson and electron mediated diagrams are suppressed respectively
by the fourth and second power of the active-sterile mixing angle $\alpha$.
Therefore we have considered other two $s$-channel diagrams only.
\begin{eqnarray}
g_{WWh}&=& \frac{2M_W^2}{v}\cos\theta\,, \nonumber \\ 
g_{WWH}&=& \frac{2M_W^2}{v}\sin\theta\,, \nonumber \\
A_{WW} &=& \frac{g_{WWh}\,g_{hN_1 N_1}
((s-M_h^2)-i\,M_h \Gamma_h)}{(s-M_h^2)^2+(M_h \Gamma_h)^2}
+ \frac{g_{WWH}\,g_{HN_1 N_1}((s-M_H^2)-i\,M_H \Gamma_H)}
{(s-M_H^2)^2+(M_H \Gamma_H)^2}\,,\nonumber \\
{\left|M_{WW}\right|^2}&=&\frac{4}{9}(s-4M_{N_1}^2)
\left(1+\frac{(s-2M_W^2)^2}{8M_W^4}\right)\left|A_{WW}\right|^2\,, \nonumber \\
\sigma_{WW} &=& \frac{1}{32\pi s}\frac{\sqrt{1-\dfrac{4M_{N_1}^2}{s}}}
{\sqrt{1-\dfrac{4M_{W}^2}{s}}}\,\,\left|M_{WW}\right|^2\,.
\label{wwn1n1}
\end{eqnarray}

\newpage

$\underline{\mathbf{Z \, Z \rightarrow N_1 \bar{N_1}}}$\\

There are two $s$-channel diagrams and two $t$-channel diagrams for
$Z \, Z \rightarrow N_1 \bar{N}_1$ annihilation process. The $t$-channel
diagrams mediated by active and sterile neutrinos are suppressed by
fourth and eighth power of $\alpha$ respectively. Hence we have considered
only two $s$-channel diagrams mediated by $h$ and $H$.
\begin{eqnarray}
g_{ZZh}&=& \frac{2M_Z^2}{v}\cos\theta\,, \nonumber \\ 
g_{ZZH}&=& \frac{2M_Z^2}{v}\sin\theta\,, \nonumber \\
A_{ZZ} &=& \frac{g_{ZZh}\,g_{hN_1 N_1}((s-M_h^2)-i\,M_h \Gamma_h)}
{(s-M_h^2)^2+(M_h \Gamma_h)^2} + \frac{g_{ZZH}\,g_{HN_1 N_1}
((s-M_H^2)-i\,M_H \Gamma_H)}{(s-M_H^2)^2+(M_H \Gamma_H)^2}\,,\nonumber \\
{\left|M_{ZZ}\right|^2}&=&\frac{4}{9}(s-4M_{N_1}^2)
\left(1+\frac{(s-2M_Z^2)^2}{8M_Z^4}\right)\left|A_{ZZ}\right|^2\,, \nonumber \\
\sigma_{ZZ} &=& \frac{1}{32\pi s}\frac{\sqrt{1-\dfrac{4M_{N_1}^2}{s}}}
{\sqrt{1-\dfrac{4M_{Z}^2}{s}}}\,\,\left|M_{ZZ}\right|^2\,.
\label{zzn1n1}
\end{eqnarray}


$\underline{\mathbf{f \, \bar{f} \rightarrow N_1 \bar{N}_1}}$ (where $f$\,denotes
any SM quarks or leptons)\\

In this annihilation process four $s$-channel diagrams, mediated by
$Z$, $\zbl$, $h$ and $H$, are possible. However, the $Z$ boson mediated
diagram is suppressed by $\alpha^4$ and consequently we have neglected it.

\begin{eqnarray}
g_{ffh}&=& - \frac{M_f}{v}\cos\theta\,, \nonumber \\ 
g_{ffH}&=& -\frac{M_f}{v}\sin\theta\,, \nonumber \\
A_{ff} &=& \frac{g_{ffh}\,g_{hN_1 N_1}((s-M_h^2)-i\,M_h \Gamma_h)}
{(s-M_h^2)^2+(M_h \Gamma_h)^2} + \frac{g_{ffH}\,g_{HN_1 N_1}
((s-M_H^2)-i\,M_H \Gamma_H)}{(s-M_H^2)^2+(M_H \Gamma_H)^2}\,,\nonumber \\
{\left|M_{ff}\right|^2}&=& \frac{g_{\rm BL}^4 q_f^2 
\left(\frac{8\,(s-4M_{N_1}^2)(s+2M_f^2)}{3\,n_c}\right)}
{(s-M_{Z_{\rm BL}}^2)^2+(\Gamma_{Z_{\rm BL}}\,M_{Z_{\rm BL}})^2}+
\frac{2}{n_c}\left|A_{ff}\right|^2(s-4M_{N_1}^2)(s-4M_f^2)\,\nonumber \\
\sigma_{ff} &=& \frac{1}{64\pi s}\frac{\sqrt{1-\dfrac{4M_{N_1}^2}{s}}}
{\sqrt{1-\dfrac{4M_{f}^2}{s}}}\,\,\left|M_{ff}\right|^2\,,
\label{ffn1n1}
\end{eqnarray}
where $n_c$ is the colour charge of the corresponding fermion ($f$).\\

Although, the annihilation processes $HH \rightarrow N_1 \bar{N}_1$,
$hh \rightarrow N_1 \bar{N}_1$ consist of two $t$-channel
and two $s$-channel diagrams, we consider only the two dominant $s$-channel
diagrams mediated by $H$ and $h$.

$\underline{\mathbf{H \, H \rightarrow N_1 \bar{N_1}}}$

\begin{eqnarray}
g_{HHh}&=& 2\,\sin\theta\cos^2\theta\,(3\lambda_2-\lambda_3)\,v_{\rm BL}
-\lambda_3\, v \cos^3\theta + 2\cos\theta\sin^2\theta(\lambda_3-3\lambda_1)
\,v+\lambda_3 \,v_{\rm BL}\sin^3\theta\,, \nonumber \\ 
g_{HHH}&=& -3\left(\lambda_3\sin\theta\cos^2\theta \,v+2\,
\cos^3\theta\lambda_2 \, v_{\rm BL}+\sin^2\theta\cos\theta
\lambda_3 \, v_{\rm BL}+2\sin^3\theta\,\lambda_1 \,v\right)\,, \nonumber \\
A_{HH} &=& \frac{g_{HHh}\,g_{hN_1 N_1}((s-M_h^2)-i\,M_h \Gamma_h)}
{(s-M_h^2)^2+(M_h \Gamma_h)^2} + \frac{g_{HHH}\,g_{HN_1 N_1}
((s-M_H^2)-i\,M_H \Gamma_H)}{(s-M_H^2)^2+(M_H \Gamma_H)^2}\,,\nonumber \\
{\left|M_{HH}\right|^2}&=&2\,(s-4M_{N_1}^2)\left|A_{ZZ}\right|^2\,, \nonumber \\
\sigma_{HH} &=& \frac{1}{32\pi s}\frac{\sqrt{1-\dfrac{4M_{N_1}^2}{s}}}
{\sqrt{1-\dfrac{4M_{H}^2}{s}}}\,\,\left|M_{HH}\right|^2\,.
\label{HHn1n1}
\end{eqnarray}


$\underline{\mathbf{h \, h \rightarrow N_1 \bar{N_1}}}$

\begin{eqnarray}
g_{hhH}&=& 2\,\sin\theta\cos^2\theta\,(\lambda_3-3\lambda_1)\,v
-\lambda_3\, v_{\rm BL} \cos^3\theta + 2\cos\theta\sin^2\theta
(\lambda_3-3\lambda_2)\,v_{\rm BL}-\lambda_3 \,v \sin^3\theta\,, \nonumber \\ 
g_{hhh}&=& 3\left(\lambda_3\sin\theta\cos^2\theta \,v_{\rm BL}-2
\,\cos^3\theta\lambda_1 \, v-\sin^2\theta\cos\theta\lambda_3 \,
v+2\sin^3\theta\,\lambda_2 \,v_{\rm BL}\right)\,, \nonumber \\
A_{hh} &=& \frac{g_{hhh}\,g_{hN_1 N_1}((s-M_h^2)-i\,M_h \Gamma_h)}
{(s-M_h^2)^2+(M_h \Gamma_h)^2} + \frac{g_{hhH}\,g_{HN_1 N_1}
((s-M_H^2)-i\,M_H \Gamma_H)}{(s-M_H^2)^2+(M_H \Gamma_H)^2}\,,\nonumber \\
{\left|M_{hh}\right|^2}&=&2\,(s-4M_{N_1}^2)\left|A_{hh}\right|^2\,, \nonumber \\
\sigma_{hh} &=& \frac{1}{32\pi s}\frac{\sqrt{1-\dfrac{4M_{N_1}^2}{s}}}
{\sqrt{1-\dfrac{4M_{h}^2}{s}}}\,\,\left|M_{hh}\right|^2\,.
\label{hhn1n1}
\end{eqnarray}
$\lambda_1,\, \lambda_2, \, \lambda_3$ has been previously
expressed in terms of the independent parameters (Eqs. (\ref{lam1}-\ref{lam3})).
\\ \\
\newpage
$\underline{\mathbf{Z_{\rm BL} \, Z_{\rm BL} \rightarrow N_1 \bar{N_1}}}$\\

This annihilation process is also mediated by two $s$-channel diagrams
and two $t$-channel diagrams. However the $t$-channel diagram mediated by
active neutrino is suppressed by $\alpha^4$. Therefore we have considered
two $s$-channel diagrams and one $t$-channel diagram mediated by $H$, $h$
and $N_1$ respectively. For simplicity, due to smallness of $\alpha$, we have
considered $\cos \alpha \simeq 1$ and consequently $\sin \alpha \simeq 0$
in the following expressions (Eq. (\ref{a1zz})-(\ref{sigzz})).
\begin{eqnarray}
A_1&=&\Bigg[-2\,g^4_{\rm BL}\Bigg(-\left(4 M_{N_1}^2-s\right){}^{3/2}
\left(4 M_{Z_{\text{BL}}}^2-s\right){}^{3/2}
\Bigg(s^2 M_{Z_{\text{BL}}}^4+20 s M_{Z_{\text{BL}}}^6-48 M_{Z_{\text{BL}}}^8+
\nonumber \\
&&2 M_{N_1}^4 \left(-56 s M_{Z_{\text{BL}}}^2+16 M_{Z_{\text{BL}}}^4+s^2\right)+
M_{N_1}^2 \left(16 s^2 M_{Z_{\text{BL}}}^2-102 s M_{Z_{\text{BL}}}^4
+184 M_{Z_{\text{BL}}}^6+s^3\right)\Bigg)
\nonumber \\&&
+12 M_{Z_{\text{BL}}}^4 \left(2 M_{Z_{\text{BL}}}^2-8 M_{N_1}^2-s\right)
\left(M_{N_1}^2 \left(s-4 M_{Z_{\text{BL}}}^2\right)+M_{Z_{\text{BL}}}^4\right)
\nonumber \\&&
2\,(s-4M_{Z_{\rm BL}}^2)(4 M_{N_1}^2-s)
\log \left(\frac{s-\sqrt{4 M_{N_1}^2-s} \sqrt{4 M_{Z_{\text{BL}}}^2-s}-
2 M_{Z_{\text{BL}}}^2}{s+\sqrt{4 M_{N_1}^2-s}
\sqrt{4 M_{Z_{\text{BL}}}^2-s}-2 M_{Z_{\text{BL}}}^2}\right)\Bigg)\Bigg]\times 
\nonumber \\&&
\dfrac{1}{27 M_{Z_{\text{BL}}}^4 \left(4 M_{N_1}^2-s\right){}^{3/2}
\left(4 M_{Z_{\text{BL}}}^2-s\right){}^{3/2} \left(M_{N_1}^2
\left(s-4 M_{Z_{\text{BL}}}^2\right)+M_{Z_{\text{BL}}}^4\right)} \,,
\label{a1zz}
\end{eqnarray}

\begin{eqnarray}
A_2&=&\Bigg[\sqrt{4 M_{N_1}^2-s} \left(2 M_{Z_{\text{BL}}}^2-s\right)
\sqrt{4 M_{Z_{\text{BL}}}^2-s} \Bigg(M_{N_1}^2
\left(88 M_{Z_{\text{BL}}}^2-46 s\right)
+s \left(20 M_{Z_{\text{BL}}}^2+s\right)\Bigg)+
\nonumber \\&&
48 s \left(M_{N_1}-M_{Z_{\text{BL}}}\right)
\left(M_{Z_{\text{BL}}}+M_{N_1}\right) \left(M_{N_1}^2
\left(4 M_{Z_{\text{BL}}}^2-s\right)-M_{Z_{\text{BL}}}^4\right) \times
\nonumber \\&&
\log \Bigg(\dfrac{1 + \frac{\sqrt{4 M_{N_1}^2-s}
\left(s-2 M_{Z_{\text{BL}}}^2\right)
\sqrt{4M_{Z_{\text{BL}}}^2-s}}{M_{N_1}^2 \left(8 M_{Z_{\text{BL}}}^2-2 s\right)
-4 s M_{Z_{\text{BL}}}^2+2 M_{Z_{\text{BL}}}^4+s^2}}{1-\frac{\sqrt{4 M_{N_1}^2-s}
\left(s-2 M_{Z_{\text{BL}}}^2\right)
\sqrt{4M_{Z_{\text{BL}}}^2-s}}{M_{N_1}^2
\left(8 M_{Z_{\text{BL}}}^2-2 s\right)
-4 s M_{Z_{\text{BL}}}^2+2 M_{Z_{\text{BL}}}^4+s^2}}\Bigg)\Bigg]
\dfrac{2\,g^4_{\rm BL}}{27 M_{Z_{\text{BL}}}^4 \sqrt{4 M_{N_1}^2-s}
\left(s-2 M_{Z_{\text{BL}}}^2\right) \sqrt{4 M_{Z_{\text{BL}}}^2-s}} \,,
\nonumber \\
\end{eqnarray}

\begin{eqnarray}
A_3&=&\Bigg[\Bigg(\sqrt{s-4 M_{N_1}^2} \sqrt{s-4 M_{Z_{\text{BL}}}^2}
\left(-2 s M_{Z_{\text{BL}}}^2+4 M_{Z_{\text{BL}}}^4+s^2\right)
\nonumber \\ &&
+2 \left(M_{N_1}^2 \left(-4 s M_{Z_{\text{BL}}}^2+8M_{Z_{\text{BL}}}^4+s^2\right)
-2 M_{Z_{\text{BL}}}^6\right)\times
\nonumber \\&&
\log \left(\frac{s-\sqrt{s-4 M_{N_1}^2} \sqrt{s-4 M_{Z_{\text{BL}}}^2}
-2 M_{Z_{\text{BL}}}^2}{s+\sqrt{s-4 M_{N_1}^2}
\sqrt{s-4 M_{Z_{\text{BL}}}^2}-2 M_{Z_{\text{BL}}}^2}\right)\Bigg) 
\nonumber \\&&
64 \, g_{\text{BL}}^4 \sin ^2\theta \,
M_{N_1}^2 \left(M_h^2-s\right)\Bigg]\times
\nonumber \\ &&
\dfrac{1}{9 M_{Z_{\text{BL}}}^4 \sqrt{s-4 M_{N_1}^2}
\sqrt{s-4 M_{Z_{\text{BL}}}^2} \left(\Gamma _h^2 M_h^2+
\left(M_h^2-s\right){}^2\right)}\,,  
\end{eqnarray}

\begin{eqnarray}
A_4&=&\Bigg[\Bigg(\sqrt{s-4 M_{N_1}^2} \sqrt{s-4 M_{Z_{\text{BL}}}^2}
\left(-2 s M_{Z_{\text{BL}}}^2+4 M_{Z_{\text{BL}}}^4+s^2\right)
\nonumber \\ &&
+2 \left(M_{N_1}^2 \left(-4 s M_{Z_{\text{BL}}}^2
+8M_{Z_{\text{BL}}}^4+s^2\right)-2 M_{Z_{\text{BL}}}^6\right)\times
\nonumber \\&&
\log \left(\frac{s-\sqrt{s-4 M_{N_1}^2} \sqrt{s-4 M_{Z_{\text{BL}}}^2}
-2 M_{Z_{\text{BL}}}^2}{s+\sqrt{s-4 M_{N_1}^2}
\sqrt{s-4 M_{Z_{\text{BL}}}^2}-2 M_{Z_{\text{BL}}}^2}\right)\Bigg) 
\nonumber \\&&
64 \, g_{\text{BL}}^4 \cos ^2\theta \, M_{N_1}^2
\left(M_H^2-s\right)\Bigg]\times
\nonumber \\ &&
\dfrac{1}{9 M_{Z_{\text{BL}}}^4 \sqrt{s-4 M_{N_1}^2}
\sqrt{s-4 M_{Z_{\text{BL}}}^2} \left(\Gamma _H^2 M_H^2+
\left(M_H^2-s\right){}^2\right)}\,,
\end{eqnarray}

\begin{eqnarray}
A_5&=&\dfrac{32 \, g_{\text{BL}}^4 \sin ^4\theta M_{N_1}^2
\left(s-4 M_{N_1}^2\right) \left(s^2-4 s M_{Z_{\text{BL}}}^2
+12 M_{Z_{\text{BL}}}^4\right)}{9\, M_{Z_{\text{BL}}}^4
\left(\Gamma_h^2 M_h^2+\left(M_h^2-s\right)^2\right)}\,,
\end{eqnarray}

\begin{eqnarray}
A_6&=&\dfrac{32 \, g_{\text{BL}}^4 \cos ^4\theta M_{N_1}^2
\left(s-4 M_{N_1}^2\right) \left(s^2-4 s M_{Z_{\text{BL}}}^2
+12 M_{Z_{\text{BL}}}^4\right)}{9\, M_{Z_{\text{BL}}}^4
\left(\Gamma_H^2 M_H^2+\left(M_H^2-s\right)^2\right)}\,,
\end{eqnarray}

\begin{eqnarray}
A_7&=&\Bigg[64\, g_{\text{BL}}^4 \sin ^2\theta\cos^2\theta M_{N_1}^2
\left(s-4 M_{N_1}^2\right) \left(s^2-4 s M_{Z_{\text{BL}}}^2
+12 M_{Z_{\text{BL}}}^4\right)\times
\nonumber \\&&
(\Gamma _h M_h \Gamma _H M_H+\left(s-M_h^2\right)
\left(s-M_H^2\right))\Bigg]\times
\nonumber \\ &&
\dfrac{1}{9 M_{Z_{\text{BL}}}^4 \left(\Gamma _h^2 M_h^2
+\left(M_h^2-s\right){}^2\right) \left(\Gamma _H^2 M_H^2
+\left(M_H^2-s\right){}^2\right)}\,,
\end{eqnarray}

\begin{eqnarray}
\sigma_{Z_{\rm BL},Z_{\rm BL}} &=& \frac{1}{32\pi s}
\frac{\sqrt{1-\dfrac{4M_{N_1}^2}{s}}}
{\sqrt{1-\dfrac{4M_{Z_{\rm BL}}^2}{s}}}\,\sum_{i=1}^{7}A_{i}
\label{zblzbln1n1}
\label{sigzz}
\end{eqnarray}
\\
\hspace{-0.5in}$\underline{\mathbf{N_{x} \, \bar{N_{x}} \rightarrow N_1 \bar{N_1}}}$
(where $N_x$\,s are the other two sterile neutrinos with $x=2,\,3$)\\

In this annihilation process we have considered the only $\zbl$ mediated
$s$-channel diagram as it contributes dominantly over the other
possible diagrams.
\begin{eqnarray}
&& g_{N_x N_x Z_{\rm BL}}= -\,g_{\rm BL}(\cos\alpha_x^2-\sin\alpha_x^2)\,,
\nonumber \\ 
&& g_{N_1 N_1 Z_{\rm BL}}= -\,g_{\rm BL}(\cos\alpha^2-\sin\alpha^2)\,,
\nonumber \\
&& \sigma_{N_x N_x} = \frac{(g_{N_x N_x Z_{\rm BL}}
\times g_{N_1 N_1 Z_{\rm BL}})^2 }{256 \, \pi\,s
\left[\left(s-M_{Z_{\text{BL}}}^2\right)^2
+\left(\Gamma_{Z_{\rm BL}}M_{Z_{\text{BL}}}\right)^2\right]}
\dfrac{\sqrt{s-4 M_{N_1}^2}}{\sqrt{s-4 M_{N_x}^2}}  
\times 
\nonumber \\ &&
\hspace{-0.5in}   
\Bigg[\frac{32 \left(4 M_{N_1}^2 \left(M_{N_x}^2 \left(-6 s M_{Z_{\text{BL}}}^2+7
   M_{Z_{\text{BL}}}^4+3 s^2\right)-s M_{Z_{\text{BL}}}^4\right)+s
   M_{Z_{\text{BL}}}^4 \left(s-4 M_{N_x}^2\right)\right)}{3
   M_{Z_{\text{BL}}}^4}\Bigg]\,. \nonumber \\  
\end{eqnarray}
Here, $\alpha_x$ is the active-sterile neutrino mixing angle of
$\nu_x$ with $N_x$ with $x=1,2,3$ and $\alpha_1$ has been denoted
simply by $\alpha$.
\newpage
\subsection{Total decay widths of $Z_{\rm BL}$, $H$ and $h$}
\label{total_decay}
The total decay width of different particles used in the
expressions for the annihilation cross section are given below :\\
\underline{\bf{Total Decay width of $Z_{\rm BL}$}}
\begin{eqnarray}
\Gamma(Z_{\rm BL} \rightarrow f \bar{f})&=&\frac{M_{Z_{\rm BL}}}
{12\, \pi}\, n_c \, (q_f \, g_{\rm BL})^2\left(1+\dfrac{2\,M_f^2}
{M_{Z_{\rm BL}}^2}\right)\, \sqrt{1-\dfrac{4\, M_f^2}{M_{Z_{\rm BL}}^2}}\,,
\nonumber \\
\Gamma(Z_{\rm BL} \rightarrow \nu_x \, \bar{\nu_x})&=&
\frac{M_{Z_{\rm BL}}}{24\pi}\,((\cos^2\alpha_x-\sin^2\alpha_x)^2
\,g_{\rm BL}^2)\left(1-\frac{4M_{\nu_{x}}^2}{M_{Z_{\rm BL}}^2}\right)^{3/2}\,,
\nonumber \\
\Gamma(Z_{\rm BL}\rightarrow N_x \, \bar{N_x})&=&
\frac{M_{Z_{\rm BL}}}{24\pi}\,((\sin^2\alpha_x-\cos^2\alpha_x)^2
\,g_{\rm BL}^2)\left(1-\frac{4M_{N_{x}}^2}{M_{Z_{\rm BL}}^2}\right)^{3/2}\,.
\end{eqnarray}
\begin{eqnarray}
\hspace{-1in}\Gamma_{Z_{\rm BL}} &= &\sum_{f}\Gamma(Z_{\rm BL}
\rightarrow f \bar{f})+\sum_{x=1}^{3}\Bigg(\Gamma(Z_{\rm BL}
\rightarrow \nu_x \, \bar{\nu_x})+\Gamma(Z_{\rm BL}
\rightarrow N_x \, \bar{N_x})\Bigg)\,.
\label{totalzbl}
\end{eqnarray}

\underline{\bf{Total Decay width of $H$}}
\begin{eqnarray}
\Gamma(H \rightarrow V \, V)&=&\dfrac{G_F\, M_H^3 \, \sin^2\theta
\, \delta V}{16\sqrt{2}\,\pi}\sqrt{1-\dfrac{4\,M_V^2}{M_H^2}}
\Bigg(1-\dfrac{4\,M_V^2}{M_H^2}+\dfrac{12\,M_V^4}{M_H^4}\Bigg)\,,
\nonumber \\
\Gamma(H \rightarrow Z_{\rm BL} \, Z_{\rm BL})&=&
\dfrac{g_{\rm BL}^2 M_H^3 \, \cos^2\theta}{8\,\pi \, M_{Z_{\rm BL}}^2}
\sqrt{1-\dfrac{4\, M_{Z_{\rm BL}}^2}{M_H^2}}
\Bigg(1-\dfrac{4\, M_{Z_{\rm BL}}^2}{M_H^2}+\dfrac{12\, M_{Z_{\rm BL}}^4}
{M_H^4}\Bigg)\,,
\label{Hzblzbl}\\
\Gamma(H \rightarrow f \, \bar{f})&=& \frac{n_c \, M_H}{8\,\pi}\,
\Bigg(\dfrac{M_f \, \sin\theta}{v}\Bigg)^2\left(1-\frac{4M_{f}^2}
{M_H^2}\right)^{3/2}\,,\nonumber \\
\Gamma(H \rightarrow h\, h)&=&\dfrac{g_{hhH}^2}
{32\,\pi\,M_H}\sqrt{1-\dfrac{4\,M_h^2}{M_H^2}}\,, \nonumber \\
\Gamma(H \rightarrow N_x \bar{N_x})&=& \frac{g_{H N_x N_x}^2\,M_H}
{16\,\pi}\left(1-\frac{4M_{N_{x}}^2}{M_{H}^2}\right)^{3/2}\,, \nonumber \\
\Gamma(H \rightarrow \nu_x \bar{\nu_x})&=& \frac{g_{H \nu_x \nu_x}^2\,M_H}
{16\,\pi}\left(1-\frac{4M_{\nu_{x}}^2}{M_{H}^2}\right)^{3/2}\,,
\end{eqnarray}
\begin{eqnarray}
\hspace{-1in}\Gamma_{H} &= &\sum_{f}\Gamma
(H \rightarrow f \bar{f})+\sum_{x=1}^{3}
\Bigg(\Gamma(H \rightarrow \nu_x \, \bar{\nu_x})
+\Gamma(H \rightarrow N_x \, \bar{N_x})\Bigg)\nonumber \\ &&
+\sum_{V=W,\,Z}\Gamma(H \rightarrow V \, V)
+\Gamma(H \rightarrow Z_{\rm BL} \,Z_{\rm BL})\,,
\label{totalH}
\end{eqnarray}
where, $\delta V=2$ for $W$ boson and 1 for $Z$ boson and
\begin{eqnarray*}
g_{H N_x N_x} =
2\,\cos\alpha_x \left(\sin\theta\sin\alpha_x 
\frac{\sqrt{M_{N_x}m_{\nu_x}}}{v}-\cos\theta\cos\alpha_x
\frac{g_{\rm BL}\,M_{N_x}}{M_{Z_{\rm BL}}}\right)\,,\nonumber \\
g_{H \nu_x \nu_x}=2\,\sin\alpha_x\left(\sin\theta\cos\alpha_x
\frac{\sqrt{M_{N_x}m_{\nu_x}}}{v}+\cos\theta\sin\alpha_x
\frac{g_{\rm BL}\,M_{N_x}}{M_{Z_{\rm BL}}}\right)\,.
\end{eqnarray*}

\underline{\bf{Total Decay width of $h$}}
\begin{eqnarray*}
\Gamma(h \rightarrow Z_{\rm BL} \, Z_{\rm BL})&=&
\dfrac{g_{\rm BL}^2 M_h^3 \, \sin^2\theta}
{8\,\pi \, M_{Z_{\rm BL}}^2}
\sqrt{1-\dfrac{4\, M_{Z_{\rm BL}}^2}{M_h^2}}
\Bigg(1-\dfrac{4\, M_{Z_{\rm BL}}^2}{M_h^2}
+\dfrac{12\, M_{Z_{\rm BL}}^4}{M_h^4}\Bigg)\,, \\
\Gamma(h \rightarrow H\, H)&=&\dfrac{g_{HHh}^2}{32\,\pi\,M_h}
\sqrt{1-\dfrac{4\,M_H^2}{M_h^2}}\,, \\
\Gamma(h \rightarrow N_x \bar{N_x})&=&
\frac{g_{h N_x N_x}^2\,M_h}{16\,\pi}
\left(1-\frac{4M_{N_{x}}^2}{M_{h}^2}\right)^{3/2}\,, \\
\Gamma(h \rightarrow \nu_x \bar{\nu_x})&=&
\frac{g_{h \nu_x \nu_x}^2\,M_h}{16\,\pi}
\left(1-\frac{4M_{\nu_{x}}^2}{M_{h}^2}\right)^{3/2}\,,
\end{eqnarray*}

\begin{eqnarray}
\hspace{-1in}\Gamma_{h} &=&\cos^2\theta\,\Gamma^{\rm SM}_h
+\Gamma(h \rightarrow Z_{\rm BL} \,Z_{\rm BL})
+\Gamma(h \rightarrow H\, H)\nonumber \\&&
+\sum_{x=1}^{3}\Bigg(\Gamma(h \rightarrow \nu_x \, \bar{\nu_x})
+\Gamma(h \rightarrow N_x \, \bar{N_x})\Bigg)\,,
\end{eqnarray}

where $\Gamma^{\rm SM}_h = 4.14 \times 10^{-3}$ MeV \cite{1107.5909}
is the decay width of SM Higgs boson with mass $125.5\,{\rm GeV}$ and 
\begin{eqnarray*}
g_{h N_x N_x}=2\,\cos\alpha_x\left(\cos\theta\sin\alpha_x
\frac{\sqrt{M_{N_x}m_{\nu_x}}}{v}+\sin\theta\cos\alpha_x
\frac{g_{\rm BL}\,M_{N_x}}{M_{Z_{\rm BL}}}\right)\,,\\
g_{h \nu_x \nu_x}=2\,\sin\alpha_x\left(\cos\theta\cos\alpha_x
\frac{\sqrt{M_{N_x}m_{\nu_x}}}{v}-\sin\theta\sin\alpha_x
\frac{g_{\rm BL}\,M_{N_x}}{M_{Z_{\rm BL}}}\right)\,.
\end{eqnarray*}
\subsection{Decay width of $N_1 \rightarrow e^\pm\,\,\nu_i$ ($i=1$ to 3)}
\label{gamman1enu}
In this section, we have calculated the three body decay width
of sterile neutrino dark matter
$N_1$ into $e^{\pm}$ and $\nu_i$. In this calculation we have considered only
$W^{\pm}$ and $Z$ bosons mediated diagrams as these three diagrams contribute
dominantly to this decay process of $N_1$. Also in our calculation,
for simplicity, we have neglected terms involving neutrino masses as these are
extremely tiny compared to the masses of other particles. Further, 
we have neglected the intergenerational mixing between active and
sterile neutrinos. We have define two quantities $X$ and $Y$ in terms of
four momentums of $\nu_i$, $e^+$ and $N_1$ as
\begin{eqnarray}
X &=& (P-p_1)^2\,, \\
Y &=& (P-p_2)^2\,,
\end{eqnarray} 
where $P$ is four momentum of $N_1$ while that of $\nu_i$ and $e^-$ are
$p_1$ and $p_2$ respectively. Now,
  
\begin{eqnarray}
B_1 &=& \dfrac{8}{M^4_Z}\Bigg(a^2_2\,\Bigg(M_Z^4
\left(4 Y M_e^2-2 M_e^4+M_{N_1}^2 (X+2 Y)-X^2-2 X Y-2
Y^2\right)\Bigg) + a_3^2\,\Bigg(M_{N_1}^2 \left(M_Z^4 \times
\right.\nonumber \\ &&\left.
(X+2 Y)-2 M_e^2 \left(-2 X M_Z^2+2 M_Z^4+X^2\right)\right)+2 M_e^2
M_{N_1}^4 \left(X-2 M_Z^2\right)+M_Z^4 \left(-\left(-4 M_e^2 \times
\right.\right.\nonumber \\ &&\left.\left.
(X+Y)+2 M_e^4+X^2+2 X Y+2Y^2\right)\right)\Bigg)\Bigg)\,\,,
\end{eqnarray}

\begin{eqnarray}
B_2 &=&-\dfrac{16}{M^4_W}\Bigg(M_e^4 \left(-M_{N_1}^2
\left(8 M_W^2+7 (X+Y)\right)+5 M_{N_1}^4+4 M_W^2
\left(M_W^2+X\right)+(X+Y)^2\right)
\nonumber \\ &&
+M_e^2 \left(M_{N_1}^2 \left(4 M_W^2
\left(M_W^2+Y\right)+(X+Y)^2\right)-M_{N_1}^4 (X+Y)-8 Y M_W^4\right)-2 M_e^6 \left(-3
M_{N_1}^2
\right. \nonumber \\ && \left.
+X+Y\right)+M_e^8+4 Y M_W^4 \left(Y-M_{N_1}^2\right)\Bigg)\,\,,
\end{eqnarray}

\begin{eqnarray}
B_3 &=&-\dfrac{16}{M^4_W}\Bigg(M_e^4 \left(Y M_{N_1}^2+M_{N_1}^4+4 M_W^2 
\left(M_W^2+X\right)+Y^2\right)+M_e^2 \left(M_{N_1}^2
\left(-4 M_W^2 (X+Y)+4 M_W^4
\right.\right. \nonumber \\ && \left. \left.
+Y^2\right)-M_{N_1}^4 \left(Y-4 M_W^2\right)-8 M_W^4
(X+Y)\right)-2 M_e^6 \left(M_{N_1}^2+Y\right)+M_e^8+4 M_W^4 (X+Y)
\nonumber \\ &&
\left(-M_{N_1}^2+X+Y\right)\Bigg)\,\,,
\end{eqnarray}

\begin{eqnarray}
B_4 &=& -\frac{8}{M_W^2 M_Z^2}\Bigg(a_2\,M_Z^2 \Bigg(M_e^4
\left(M_{N_1}^2+2 M_W^2-X-2 Y\right)+M_e^2
\left(-M_{N_1}^2 (3 X+2 Y)+2
M_{N_1}^4+2 M_W^2 \times
\right. \nonumber \\ && \left.
(X-2 Y)+(X+Y)^2\right)
+M_e^6+2 Y M_W^2 \left(Y-M_{N_1}^2\right)\Bigg)
+ a_3\,
\Bigg(M_{N_1}^2 \left(M_Z^2 \left(-M_e^2
\left(4 M_W^2+3 X
\right.\right.\right. \nonumber \\ && \left.\left.\left.
+4 Y\right)+5 M_e^4+2 Y M_W^2\right)
+X M_e^2 \left(-2 M_e^2+2 M_W^2+X+Y\right)\right)
+M_e^2 M_{N_1}^4 \left(M_e^2-2 M_W^2+2M_Z^2
\right. \nonumber \\ && \left.
-X\right)
+M_Z^2 \left(-2 M_W^2 \left(-M_e^2 (X+2 Y)+M_e^4+Y^2\right)-M_e^4 (3 X+2Y)
+M_e^2 (X+Y)^2+M_e^6\right)
\Bigg)\Bigg)\,\,,
\nonumber \\
\end{eqnarray}
   
\begin{eqnarray}
B_5 &=&  -\frac{8}{M_W^2 M_Z^2}\Bigg(a_2\,M_Z^2 \Bigg(M_e^2
\left(-X M_{N_1}^2+M_{N_1}^4-2 M_W^2 (X+2 Y)+Y^2\right)
+M_e^4\left(-M_{N_1}^2+2 M_W^2+X
\right. \nonumber \\ && \left.
-2 Y\right)
+M_e^6+2 M_W^2 (X+Y)
\left(-M_{N_1}^2+X+Y\right)\Bigg)
+ a_3\,
\Bigg(
M_e^2 \left(M_{N_1}^2 \left(M_Z^2
\left(-4 M_W^2+X+2 Y\right)
\right.\right. \nonumber \\ && \left. \left.
+2 X M_W^2-X Y\right)
+M_{N_1}^4\left(-\left(2 M_W^2+M_Z^2\right)\right)
+M_Z^2 \left(M_W^2 (6 X+4 Y)+Y^2\right)\right)
+M_e^4\left(M_{N_1}^4-M_Z^2 \times
\right. \nonumber \\ && \left.
\left(M_{N_1}^2+2 M_W^2+X+2 Y\right)\right)
+M_e^6 M_Z^2-2 M_W^2 M_Z^2(X+Y)
\left(-M_{N_1}^2+X+Y\right)
\Bigg)
\Bigg)\,\,,\nonumber \\
\end{eqnarray}

Also,

\begin{eqnarray}
f_1 &=& \Bigg(\dfrac{e\,\sin 2\,\alpha}{2\,\sin 2\,
{\theta_{\rm W}}}\Bigg)^2\,,\nonumber \\
f_2 &=& \Bigg(\dfrac{e^2 \sin 2\,\alpha}{16\,\sin^2 \theta_{\rm W}}
\Bigg)^2\,, \nonumber \\
f_3 &=& -\dfrac{e^3\,\sin^2 2\,\alpha}{32\,\sin^2
\theta_{\rm W}\,\sin 2\,\theta_{\rm W}}\,, \nonumber \\
\end{eqnarray}

and

\begin{eqnarray}
D_1 &=& (X-M^2_Z)^2 + (\Gamma_Z\,M_Z)^2\,, \nonumber \\
D_2 &=& (M^2_{N_1} + M^2_{\nu_1} + 2\,M^2_e-X-Y-M^2_W)^2
+ (\Gamma_W\,M_W)^2\,,\nonumber \\
D_3 &=& (Y-M^2_W)^2 + (\Gamma_W\,M_W)^2\,, \nonumber \\
D_4 &=& D_1 D_2 \,,\nonumber \\
D_5 &=& D_1 D_3 \,,\nonumber \\
\end{eqnarray}
with $\Gamma_W$ and $\Gamma_Z$ are the total
decay widths of $W^\pm$ and $Z$ bosons respectively.

Therefore,

\begin{eqnarray}
|M^2_1| &=& \dfrac{1}{2}\,\dfrac{B_1 f_1}{D_1} \times \left(
\sum_{\alpha = 1}^{3} {U^{{}^2}_{e\,\alpha}} \right)
\,, \nonumber \\
|M^2_2| &=& \dfrac{1}{2}\,\dfrac{B_2 f_2}{D_2}\times \left(
\sum_{\alpha = 1}^{3} {U^{{}^4}_{e\,\alpha}} \right)
\,, \nonumber \\
|M^2_3| &=& \dfrac{1}{2}\,\dfrac{B_3 f_2}{D_3}
\times \left(
\sum_{\alpha = 1}^{3} {U^{{}^4}_{e\,\alpha}} \right)
\,, \nonumber \\
|M^2_4| &=& \dfrac{B_4 f_3 \left[(M^2_{N_1} + M^2_{\nu_1} +
2\,M^2_e-X-Y-M^2_W)(X-M^2_Z) + \Gamma_W\,\Gamma_Z\,M_W\,M_Z \right]}
{D_4}\times \left(
\sum_{\alpha = 1}^{3} {U^{{}^3}_{e\,\alpha}} \right)
\,, \nonumber \\
|M^2_5| &=& \dfrac{B_5 f_3 \left[(Y-M^2_W)(X-M^2_Z) +
\Gamma_W\,\Gamma_Z\,M_W\,M_Z \right]}
{D_5}\times \left(
\sum_{\alpha = 1}^{3} {U^{{}^3}_{e\,\alpha}} \right)
\,, \nonumber \\
\end{eqnarray}
where $U_{e\,\alpha}$s are the elements of PMNS matrix of neutrino mixing and
in terms of neutrino mixing angles these are defined as (assuming Dirac CP phase
$\delta = 0$)
\begin{eqnarray}
U_{e\,1} = \cos \theta_{12}\,\cos \theta_{13}\,,\,\,\,
U_{e\,2} = \sin \theta_{12}\,\cos \theta_{13}\,,\,\,\,
U_{e\,3} = \sin \theta_{13}\,.
\end{eqnarray} 
 
Finally, the expression of Matrix amplitude square for the process
$N_1\rightarrow e^\pm\,\nu_i$ is given by,
\begin{eqnarray}
|M^2| = \sum_{1}^{5}\,|M_i^2|\,. 
\end{eqnarray}

The corresponding decay width in terms of $|M^2|$ is given by,
\begin{eqnarray}
\Gamma_{N_1 \rightarrow e^{\pm} \nu_{i}} = \dfrac{1}{2\pi^3}
\dfrac{1}{32\,M^3_{N_1}}\int^{Y_{max}}_{Y_{min}}
\int^{X_{max}}_{X_{min}} |M^2|\,dX\, dY \,.\
\label{n1enu}
\end{eqnarray}
The upper and lower limits of the quantities $X$ and $Y$ are given below
\begin{eqnarray}
X_{min} &=& (x+y)^2 - \left(\sqrt{x^2-M^2_e}-\sqrt{y^2-M^2_e}\right)^2\,,\nonumber \\
X_{max} &=& (x+y)^2 - \left(\sqrt{x^2-M^2_e}+\sqrt{y^2-M^2_e}\right)^2\,,\nonumber \\
\end{eqnarray}
with
\begin{eqnarray}
x &=&\dfrac{Y + M^2_e}{2\,\sqrt{Y}}\,,\nonumber \\
y &=&\dfrac{\mdm^2 - Y - M^2_e}{2\,\sqrt{Y}}\,,\nonumber \\
\end{eqnarray}  
and 
\begin{eqnarray}
Y_{min} &=& M^2_e \,,\nonumber \\
Y_{max} &=& (\mdm-M_e)^2 \,.
\end{eqnarray}

\end{document}